\documentclass[journal]{IEEEtran}
\ifCLASSINFOpdf

\else

\fi
\usepackage{amsmath}
\usepackage{makeidx}  
\usepackage{graphicx}
\usepackage{subfigure}
\usepackage{epstopdf}
\usepackage{bm}
\usepackage{cite}
\usepackage{stfloats}
\usepackage{multirow}   

\usepackage{float}
\usepackage{booktabs}
\usepackage{algorithm}
\usepackage{algorithmic}

\usepackage{amssymb}
\usepackage{graphicx}
\usepackage{url}

\usepackage{tabularx,booktabs}
\newcolumntype{C}{>{\centering\arraybackslash}X} 
\usepackage{lipsum}

\usepackage{makecell} 

\usepackage{graphicx}
\usepackage{color}

\begin{document}

\title{Beamforming Design and Association Scheme \\
for Multi-RIS Multi-User mmWave \\
Systems Through Graph Neural Networks
}
\author{\IEEEauthorblockN{Mengbing~Liu, \textit{Graduate Student Member, IEEE}, Chongwen~Huang, \textit{Member, IEEE}, \\Ahmed Alhammadi, \textit{Member, IEEE}, 
Marco Di Renzo, \textit{Fellow, IEEE},  \\ M\'{e}rouane Debbah, \textit{Fellow, IEEE},  
 and Chau Yuen, \textit{Fellow, IEEE} }\\
 \thanks{Mengbing Liu and Chau Yuen are with the School of Electrical and Electronics Engineering, Nanyang Technological University, Singapore 639798 (E-mails: {mengbing001@e.ntu.edu.sg}; {chau.yuen@ntu.edu.sg}).}
\thanks{ Chongwen Huang is with College of Information Science and Electronic Engineering, Zhejiang University, Hangzhou 310027, China, and Zhejiang Provincial Key Laboratory of Multi-Modal Communication Networks and Intelligent Information Processing, Hangzhou 310027, China (E-mail: {chongwenhuang@zju.edu.cn}).}
\thanks{Ahmed Alhammadi is with the Technology Innovation Institute, Masdar City, Abu Dhabi, United Arab Emirates (E-mail: {{Ahmed.Alhammadi@tii.ae}}).}
\thanks{Marco Di Renzo is with Universit\'e Paris-Saclay, CNRS, CentraleSup\'elec, Laboratoire des Signaux et Syst\`emes, 3 Rue Joliot-Curie, 91192 Gif-sur-Yvette, France.
(E-mail: { marco.di-renzo@universite-paris-saclay.fr}), and with King's College London, Centre for Telecommunications Research -- Department of Engineering, WC2R 2LS London, United Kingdom (E-mail: {marco.di$\_$renzo@kcl.ac.uk}).
}
\thanks{M\'{e}rouane Debbah is with  KU 6G Research Center, Department of Computer and Information Engineering, Khalifa University, Abu Dhabi 127788, UAE and also with CentraleSupelec, University Paris-Saclay, 91192 Gif-sur-Yvette, France (E-mail: {merouane.debbah@ku.ac.ae}).}
\thanks{Parts of this paper have been presented in part at the IEEE International Conference on Communication (ICC), 2023 \cite{liu2023cooperative}.} 
}
\maketitle
\begin{abstract}

 Reconfigurable intelligent surface (RIS) is emerging as a promising technology for next-generation wireless communication networks, offering a variety of merits such as the ability to tailor the communication environment. 
 Moreover, deploying multiple RISs helps mitigate severe signal blocking between the base station (BS) and users, providing a practical and efficient solution to enhance the service coverage.
 However, fully reaping the potential of a multi-RIS aided communication system requires solving a non-convex optimization problem. This challenge motivates the adoption of learning-based methods for determining the optimal policy.
In this paper, we introduce a novel heterogeneous graph neural network (GNN) to effectively leverage the graph topology of a wireless communication environment.
  Specifically, we design an association scheme that selects a suitable RIS for each user. 
  Then, we maximize the weighted sum rate (WSR) of all the users by iteratively optimizing the RIS association scheme, and beamforming designs until the considered heterogeneous GNN converges. Based on the proposed approach, each user is associated with the best RIS,  which is shown to significantly improve the system capacity in multi-RIS multi-user millimeter wave (mmWave) communications.
Specifically, simulation results demonstrate that the proposed heterogeneous GNN closely approaches the performance of the high-complexity alternating optimization (AO) algorithm in the considered multi-RIS aided communication system, and it outperforms other benchmark schemes. Moreover, the performance improvement achieved through the RIS association scheme is shown to be of the order of $30\%$.
\end{abstract}


\IEEEpeerreviewmaketitle

\section{Introduction}

Millimeter wave (mmWave) communications play an indispensable role in providing high-bandwidth and high-throughput services for multimedia applications in current and next-generation wireless systems \cite{xiao2017millimeter}.
Although mmWave communications allow for higher data rates due to the wider bandwidth, signals transmitted in high-frequency bands experience serious propagation attenuation and penetration losses \cite{tataria20216g}. 
As a remedy, reconfigurable intelligent surface (RIS) has emerged as a promising technology to enhance the link quality of service and restore communication in dead zones. Compared to alternatives like ultra-dense networks and massive active antenna arrays, RIS is an effective and cost-efficient paradigm for assisting mmWave networks through a large number of nearly passive reflective elements \cite{huang2019reconfigurable,direnzo2020smart,huang2021multi,yang2021intelligent}. 

The integration of RIS into various communication systems may improve the network capacity and reduce the energy consumption. 
 However, jointly designing the beamforming of base stations (BSs) and RISs remains a critical challenge, as the corresponding optimization problem for distributed RIS-aided systems is non-convex and computationally intensive.
 Consequently, significant efforts have been directed toward developing efficient algorithms to address this challenge in RIS-aided systems \cite{ning2020beamforming,cao2020intelligent,chen2022IRS,gong2023holographic,an2023stacked,shi2024ris, Basar2024reconfigurable}.  
Recently, the beamforming design in a multi-RIS aided system has been investigated to compensate for the severe propagation attenuation in high-frequency bands \cite{cao2021reconfigurable, Nguyen2023leveraging, Abrardo2021Intelligent, Abrardo2021MIMO, Perovi2022On}. 
These problem-specific algorithms, however, are laborious and require extensive knowledge, particularly in multi-RIS aided systems.

Inspired by the success of deep learning in various application domains, e.g., computer vision, deep neural network (DNN)  models have been employed in wireless communication systems in recent years \cite{ WEI2023, huang2024challenges, gao2020unsupervised,xu2021robust,alexandropoulos2022pervasive,cao2021ai}. For instance, an unsupervised deep learning approach was proposed in \cite{gao2020unsupervised} to obtain the optimal phase configuration in an RIS-aided single-user wireless communication system.
In \cite{xu2021robust},  a deep quantization neural network was developed for optimizing the beamforming matrices at the BS and RIS, by assuming discrete phase shifts and imperfect channel state information (CSI).

However, such DNN models may not be well-suited for applications in wireless communications, as they fail to capture the underlying network topology, often leading to dramatic performance degradation when the network size increases.  
To enhance the scalability and generalization of deep learning for wireless applications,  it is crucial to incorporate the network structure into the neural network architecture \cite{bengio2021machine}. 

From this perspective, graph neural networks (GNNs) have emerged as a suitable model to exploit the inherent graph topology of wireless communication networks\cite{he2021overview, shen2022graph, he2022gblinks}. 
A GNN is a type of neural network designed to operate on graphs and other structured data, allowing for the utilization of abundant information with structural relations among the network nodes \cite{schlichtkrull2018modeling}. 
For example, recent research works apply GNNs for resource allocation in device-to-device (D2D) communication networks \cite{shen2020graph,zhang2021scalable}. In these works, the service links and interference relations are modeled as nodes and edges, respectively.
Based on these graph embedding features, the beamforming design for maximizing the weighted sum rate (WSR) can be solved efficiently. Recently, GNN architectures have been applied to the design of RIS-aided systems\cite{jiang2021learning,zhang2022learning,chan2024beamforming,yeh2024enhanced}.
In \cite{jiang2021learning}, a GNN was employed to learn both BS and RIS beamforming directly from received pilots, thereby maximizing the WSR without explicit channel estimation. Building on this GNN architecture, \cite{zhang2022learning} proposed a three-stage algorithm with two GNNs to jointly determine beamforming matrices and user scheduling when the number of users greatly exceeds the number of BS antennas. {
Recently, in \cite{chan2024beamforming}, a GNN-based framework jointly optimizes user association, BS beamforming, and RIS phase adjustment without explicit CSI, boosting sum rate and enhancing load balancing. Meanwhile, \cite{yeh2024enhanced} proposes an enhanced-GNN (E-GNN) featuring improved angular feature extraction, graph attention, and transfer learning to overcome low signal-to-noise ratio (SNR) instability and multi-path challenges, achieving near-optimal sum-rate performance.
}
 
In this paper, we leverage the advantages of GNNs for beamforming optimization in a multi-RIS multi-user mmWave communication system.
{Besides beamforming optimization, we consider a more general problem in which each user selects the most suitable RIS to improve the link quality of service and energy efficiency.
 In single RIS systems, the BS must spread its power via one RIS to serve all the users, which is not efficient. By assigning each user to a specific RIS, instead, the system can adjust the RIS phase configurations to concentrate the signal on each user, reducing power waste. Furthermore, the RIS association scheme improves system flexibility and scalability while reducing the computational overhead.}
 
For this reason, an RIS association design is introduced into the beamforming design problem to select the most suitable RIS for each user so as to maximize the WSR, i.e., aiming to provide the best performance while reducing the number of RISs.
{
Though there exist some research works analyzing the BS-RIS-user association problem {\cite{zhu2022drl,he2023joint,huang2022empowering}}, they focus on the association between multiple BSs and users in  RIS-aided and full-connected multi-RIS aided communication scenarios. 
 Different from previous works, we primarily consider the association scheme between multiple RISs and multiple users to improve the system's performance while improving the flexibility and efficiency of the system by selecting the most appropriate RIS.
}

 Meanwhile, existing GNN structures with single-type nodes have only been applied to the beamforming design in RIS-aided communication systems, and do not incorporate the association scheme between multiple RISs and multiple users \cite{shen2020graph,zhang2021scalable}. Considering the strong coupling between the beamforming and association designs for multiple RISs, we propose a new graph modeling approach to better map the beamforming design and RIS association scheme in the proposed system, i.e., a heterogeneous GNN, which contains multi-type nodes with specific node features tailored to the system requirements \cite{zhang2019heterogeneous}. Overall, the main contributions of this work are the following:

\begin{itemize}
\item We first present a heterogeneous GNN structure and map it onto an optimization problem tailored to multi-RIS aided systems.
The BS, RISs, and users are represented as three different types of nodes and we embed different optimization variants as features of the corresponding types of nodes, i.e., the beamforming matrix in the BS node, the phase configuration matrix in each RIS node, and the RIS association matrix in the user node, respectively.

\item Two different loss functions are adopted to address the challenge of learning continuous (the beamforming matrix and phase configuration matrices) and discrete variables (the RIS association matrix) simultaneously. One is the WSR of the multi-RIS aided communication system, and maximizing it is the target of the considered optimization problem. 
 The second one is the categorical cross-entry (CE) loss, which helps the learning process of the RIS association scheme.  
 In order to balance the significance between two loss functions during training, a penalty term is utilized for the loss function. 

\item To thoroughly verify the effectiveness of the proposed structure, a series of commonly used benchmarks are illustrated.
The optimized results are very close to those obtained by using {the traditional alternating optimization (AO) algorithm}. In addition, it is shown that the proposed heterogeneous GNN scheme performs $10$ times better than a DNN-based scheme.

\item We conduct a comparative experiment, where the proposed RIS  association scheme is determined in advance according to the distance.
Then, the heterogeneous GNN network is used to learn the optimal beamforming and phase configuration matrices. Extensive experiments demonstrate that the WSR obtained by jointly optimizing the three variables is higher than that of using the predetermined scheme. Finally, compared to the case study when the users are associated with the  single RIS, the performance improvement achieved by the proposed association scheme is around $30\%$. 
\end{itemize}

The remainder of this paper is organized as follows:
Section II describes the multi-RIS aided system model and the channel model for mmWave bands. Also, the problem formulation for maximizing the WSR is introduced.
Section III presents the message-passing paradigm and normalization layers for the proposed heterogeneous GNN scheme. 
Section IV illustrates the complete algorithmic frameworks with different predetermined RIS association strategies.
Section V reports the simulation results of the proposed GNNs and the comparison with relevant benchmarks verifying the effectiveness of the proposed approach. Finally, Section VI concludes the article.

\textit{Notation}: Uppercase (lowercase) boldface letters denote matrix (vector)  and the field of complex (real) numbers is denoted by $\mathbb{C} (\mathbb{R})$. $(\cdot)^T$ and $(\cdot)^H $ indicate the operations of transpose, and Hermitian transpose.
 $\left[ \cdot \right]_{m,n}$ denotes the elements at the $m$-th row and the $n$-th column of a matrix, and  
${\rm{diag}}({\mathbf{x}})$ is a diagonal matrix whose diagonal elements are formed with the elements of $\mathbf{x}$. 
Besides, $\mathbf{A} \otimes \mathbf{B}$ denotes the Kronecker product of $\mathbf{A}$ and $\mathbf{B}$.
For a complex number, $\operatorname{Re} \{ \cdot\}$  and  $\operatorname{Im}\{ \cdot \}$ denote the real part
and imaginary part. 
$|\cdot|$ and $ \|\cdot\|_{\mathcal{F}}$ represent the absolute value and Frobenius norm, respectively. $v \sim  \mathcal{CN}(0,\sigma^2)$ means that $v$ follows the complex Gaussian distribution with zero-mean and variance   $\sigma^2$.

\section{Multi-RIS Multi-user Mmwave Communication Networks}

In this section, we introduce a multi-RIS aided mmWave communication system model and the RIS association scheme. Also, the channel model for mmWave systems and the WSR maximization problem with respect to the beamforming design and the RIS association scheme are presented.

\subsection{System Model}

We consider a multi-RIS aided system with  an $N_t$-antenna BS, a set of single-antenna users $\mathcal{K} = \left\{1,2,\cdots,K \right\} $ and a set of RISs $\mathcal{R} = \left\{1,2,\cdots, R \right\}$. Each RIS is equipped with $M = M_x \times M_y$ reflecting elements, where $M_x$ and $M_y$ represent the number of horizontal and vertical elements, respectively. Assume that the BS transmits $K$ data streams to the users through beamforming optimization. Thus, the transmitted signal $\mathbf{x}$ from the BS can be expressed as
\begin{align}
\mathbf{x} 
   &\triangleq \mathbf{F} \mathbf{s} 
   = \sum_{k = 1}^{K} \mathbf{f}_k {s}_{k},
\end{align}
where $\mathbf{s} = \left[ s_1,\cdots,s_K\right]^{T} \in \mathbb{C}^{K \times 1}$ indicates the symbol vector and $\mathbb{E}({\mathbf{s}\mathbf{s}^H}) = \mathbf{I}_K$.  $\mathbf{F} = \left[\mathbf{f}_{1}, \cdots, \mathbf{f}_{K} \right]  \in \mathbb{C}^{N_\mathrm{t} \times K}$ denotes the beamforming matrix at the BS and the power constraint is denoted by  $\left\|\mathbf{F}\right\|_\mathcal{F}^2 \leq P_{{\max}}$, where $P_{{\max}}$ is the maximum transmit power at the BS.

The direct BS-user links are assumed to be blocked by obstacles. The BS can only communicate with the users via the reflections from the RISs. Therefore, it is crucial to investigate the joint design of RIS  association, BS beamforming, and the phase configurations of the RISs. In addition, we assume that the CSI of all the links is perfectly known to the BS\footnote{
Several research papers can be found in the literature with focus on channel estimation for RIS-aided channels \cite{Ma2021Model,  He2022learning,liu2022deep,wei2021channel,guo2022uplink,pan2022An,zhi2023two}. Therefore, we do not focus on these aspects in the present paper.}, and all the channels experience quasi-static flat fading. 

 \begin{figure}[ht]
\setlength{\abovecaptionskip}{0pt}
\setlength{\belowcaptionskip}{0pt}
\centering
\includegraphics[width= 0.5\textwidth]{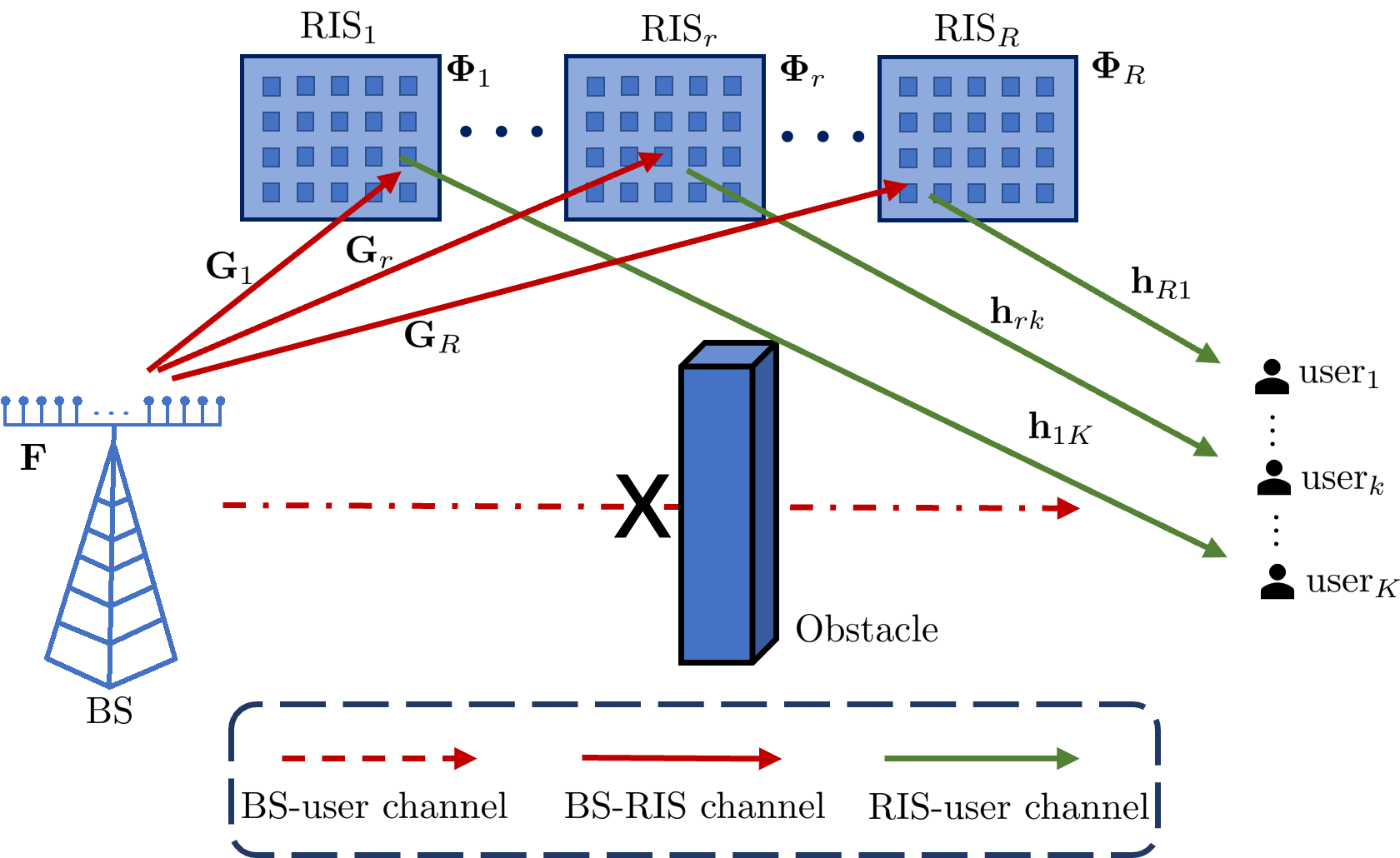}
\DeclareGraphicsExtensions.
\caption{Illustration of the multi-RIS aided system model.}
\vspace{-5pt}
\end{figure}

In this paper, each user is associated with exactly one RIS, which is selected from the available RISs to optimize the communication performance by maximizing the WSR of the multi-RIS aided communication system. To easily express the association between the RISs and users, a matrix $\mathbf{U} = [\mathbf{u}_1, \cdots, \mathbf{u}_K]^{T} \in \mathbb{R}^{K \times R}$ consisting of binary variables $u_{k,i}$ is defined as  
\begin{equation}
   u_{k,i}=\left\{\begin{array}{l}
1,  \text {user}_k \text { is assisted by RIS}_i \\
0, \text { otherwise. }
\end{array}\right.
\end{equation}

Based on the RIS association scheme, the equivalent channel $\mathbf{h}_{k}$ from the BS to user$_k$, including the cascaded links, can be expressed as
\begin{equation}
\mathbf{h}_{k} 
   =\sum_{i \in \mathcal{R}} u_{k, i} \mathbf{h}_{ik} {\mathbf{\Phi}}_{i}\mathbf{G}_{i},
\end{equation}
where $\mathbf{G}_{i} \in \mathbb{C}^{M \times N_t}$ and  $\mathbf{h}_{ik} \in \mathbb{C}^{1\times M}$ denote the BS-RIS$_i$ and RIS$_i$-user$_k$ channel, respectively. $\boldsymbol{\Phi}_{i} = {\rm{diag}} \left( \theta_{i,1},\cdots,\theta_{i,M} \right) \in \mathbb{C}^{M \times M}$ with $\theta_{i,m} = e^{j\phi_{i,m}} $ denotes the phase shift matrix of RIS$_i$. 

{In the considered model, the RIS association scheme optimizes the signal reflections by matching users with specific RISs, focusing on enhancing the signal quality for the associated users.  {Considering the distributed arrangement of RISs and the consequent signal attenuation due to propagation distances, the signals reflected toward non-associated users are markedly attenuated compared to those directed at associated users.} Therefore,  some interfering signals, for ease of analysis, are not considered due to their weak impact. Thus, the signal $\mathbf{y}_{k}$ received by the user$_k$ can be rewritten as } 
\begin{align}
\mathbf{y}_{k} &=  \mathbf {h}_{k} \mathbf{x} + \mathbf{n}_{k} \\\nonumber
&= \left( \sum_{i \in \mathcal{R}} u_{k,i} \mathbf{h}_{ik} {\mathbf{\Phi}}_{i}\mathbf{G}_{i}\right)\mathbf{f}_{k}{s}_{k}\\\nonumber
&+\sum_{j\in \mathcal{K}, j \neq k}^{K} \left( \sum_{i \in \mathcal{R}} u_{k,i} \mathbf{h}_{ik} {\mathbf{\Phi}}_{i}\mathbf{G}_{i}\right)  \mathbf{f}_{j} {s}_{j} +  \mathbf{n}_k.
\end{align}

According to (4), the signal-to-interference-plus-noise-ratio (SINR) at user$_k$ is given as
\begin{equation}
{\rm{SINR}}_{k} \!\!= \frac{\left| \left( \sum\limits_{r \in \mathcal{R}} u_{r,k} \mathbf{h}_{rk} {\mathbf{\Phi}}_{r}\mathbf{G}_{r}   \right)  \mathbf{f}_{k}\right|^{2}}{ \sum\limits_{j\in \mathcal{K}, j \neq k}^{K} \left|
\left( \sum\limits_{r \in \mathcal{R}} u_{r,k} \mathbf{h}_{rk} {\mathbf{\Phi}}_{r}\mathbf{G}_{r}  \right)  \mathbf{f}_{j}
\right|^{2} \!\!\!+\! \sigma^{2}},
\end{equation}
where $\sigma^{2}$ denotes the effective noise variance.

\subsection{Channel Model}

In this paper, we consider the typical Saleh-Valenzuela mmWave channel model\cite{akdeniz2014millimeter} with a limited number of scattering paths to characterize the sparse nature of mmWave channels. We assume that the BS is equipped with a uniform linear array (ULA) whose array response vector is given as 
\begin{align}
\mathbf{a}_{\mathrm{ULA}}(\phi) &= \frac{1}{\sqrt{N}}\left[e^{-j \frac{2 \pi d}{\lambda} \phi_i}\right]_{i \in \mathcal{I}(N)},  
\end{align}
\begin{align}
\mathcal{I}(N)=\{n-(N-1) / 2, n=0,1, \cdots, N-1\},
\end{align}
where $\lambda$ and $d$ represent the signal wavelength and the antenna spacing, respectively. Besides, the number of antennas is $N$  and $\phi$ denotes the angle of arrival (AoA) or the angle of departure (AoD).

Each RIS is modeled as a uniform planar array (UPA) with $M = M_y \times M_x$ elements. The corresponding array response vectors are given by
\begin{align}
 \mathbf{a}_{\mathrm{UPA}}(\phi^1,\phi^2) 
 &= \mathbf{a}_{\mathrm{ULA}}^{\mathrm{x}}(\phi^1) \otimes \mathbf{a}_{\mathrm{ULA}}^{\mathrm{y}}(\phi^2),
\end{align}
where $\phi^1$ and $\phi^2$ denote the azimuth and elevation angles, respectively, and $\mathbf{a}_{\mathrm{ULA}}^{\mathrm{x}}(\phi^1)$ and $\mathbf{a}_{\mathrm{ULA}}^{\mathrm{y}}(\phi^2)$ are defined as $\mathbf{a}_{\mathrm{ULA}}(\phi)$. The array element spacings of the considered ULA and UPA are assumed to be $\lambda / 2$. Due to the severe path loss, the transmit power of two or more reflections can be ignored \cite{wei2022joint}. Hence, according to the considered array response vectors, the effective channel $\mathbf{h}_{ik}$ for the RIS$_i$-user$_k$  and $\mathbf{G}_{i}$ for the BS-RIS$_i$ can be represented as 
\begin{align}
   \mathbf{h}_{ik} &= \sum_{l_h=1}^{L_h}\beta_{l_h} {\mathbf{a}}_{\mathrm{r}}^H \left(\phi_{l_h}^{{\mathrm{r}},1}, \phi_{l_h}^{{\mathrm{r}},2} \right), \\
    \mathbf{G}_{i} &= \sum_{l_g=1}^{L_g}\beta_{l_g}{\mathbf{a}}_{\mathrm{r}} \left(\phi_{l_g}^{{\mathrm{r}},1}, \phi_{l_g}^{{\mathrm{r}},2} \right) {\mathbf{a}}_{
    \mathrm{b}}^H \left(\phi_{l_g}^{\mathrm{b}} \right),
\end{align}
where $L_g$ ($L_h$) denotes the number of paths, including one line-of-sight (LoS) path and $ L_g(L_h)-1 $ non-line-of-sight (NLoS) paths for the cascaded links $\mathbf{G}_{i}$ ($\mathbf{h}_{ik}$). Similarly, $\beta_{l_g}$  ($\beta_{l_h}$) denotes the complex path gain of the $l_{g}$ ($l_{h}$) path of the corresponding channels. Note that mmWave channels normally consist of only a small number of dominant multi-path components \cite{alkhateeb2014channel,wang2020intelligent}, while the scattering at sub-6 GHz is generally rich. Here, $\phi_{l_g}^{\mathrm{b}}$ denotes the AoD at the BS while $\phi_{l_h}^{\mathrm{r},2}$ and $\phi_{l_g}^{\mathrm{r},2}$   {  denote  the elevation angles of the cascaded links $\mathbf{h}_{ik}$ and $\mathbf{G}_{i}$, respectively.}
 In addition, $\phi_{l_h}^{\mathrm{r},1}$ and $\phi_{l_g}^{\mathrm{r},1}$ represent the azimuth of the cascaded links $\mathbf{h}_{ik}$ and $\mathbf{G}_{i}$, respectively.

\subsection{Problem Formulation}

Our main objective is to maximize the WSR by designing the beamforming matrix $\mathbf{F}$, the RIS  association matrix $\mathbf{U}$ as well as the beamforming matrices $\mathbf{\Phi}_i, \forall i \in \mathcal{R}$. First, the WSR of the multi-RIS aided system is given by
{ 
\begin{align}
    W_R &= \sum_{k = 1}^{K} w_k W_{R_k} \\\nonumber
    &= \sum_{k = 1}^{K} w_k\rm{log}_2(1+ {\rm{SINR}}_k),
\end{align}}
where $w_k > 0$ denotes the weight for user$_k$ with $\sum_{k = 1}^K w_k = 1 $. 
Subsequently, the WSR maximization problem can be formulated as
\begin{subequations}
\begin{align}
\text { P1: }
{\mathop{\max}\limits_{\mathbf{F}, \mathbf{U}, \mathbf{\Phi}_i,\forall i \in \mathcal{R} }} & \sum_{k = 1}^{K}w_k \rm{log}_2(1+ {\rm{SINR}}_k) \\
\text { s.t. } 
& \left\|\mathbf{F} \right\|_\mathcal{F}^{2} \leq P_{ {\max}} \\
& \left|\boldsymbol{\Phi}_{i,m}\right| = 1,  \forall m = 1,2, \cdots, M \\
& \sum_{i \in \mathcal{R}} u_{k,i} = { 1}, \forall k \in \mathcal{K} \\
& u_{k,i} \in \left\{0,1\right\}, \forall k \in \mathcal{K}, \forall i \in \mathcal{R}.
\label{P1}
\end{align}
\end{subequations}
where (12a) denotes the optimization objective for maximizing the WSR, and achieving this goal depends on the beamforming matrix $\mathbf{F}$, phase configuration matrices $\mathbf{\Phi}_i, \forall i \in \mathcal{R}$ and the RIS  association matrix $\mathbf{U}$. The constraint (12b) imposes that the maximum transmit power of the BS is $P_{ {\max}}$, while the constraint (12c) ensures that the element in each RIS satisfies the unit-modulus requirement. In addition, the constraints (12d) and (12e) are related to the RIS association scheme, where (12d) requires each user to choose the { most appropriate} RIS as the cascaded link, and (12e) accounts for the binary nature of the element $u_{k,i}$ in $\mathbf{U}$.

Solving problem (12) is computationally challenging because it is a non-convex problem involving both discrete variables (the RIS association matrix) and continuous variables (the reflection coefficients at the RISs and the beamforming matrix at the BS). 
In addition, it is difficult to decouple the optimization variables as they are closely interrelated. 
To tackle this problem, a GNN model is utilized as a powerful function approximator \cite{shen2022graph} to learn the transmission scheme from the CSI data. The structure of the proposed heterogeneous GNN will be introduced in detail in the next section. 

\section{Weighted Sum Rate Maximization via Heterogeneous Graphs Neural Networks}

In order to solve the rate maximization problem P1 via a heterogeneous GNN, we need to map the problem onto a GNN. Then, a message-passing paradigm is utilized to learn the optimal policy based on the channel state information. 

\subsection{Graph Modeling of Multi-RIS Aided Networks}

To solve P1, we model the considered network as a graph to solve P1. A basic GNN can be formally characterized by a tuple $\mathcal{G}=$ $(\mathcal{V}, \mathcal{E})$, where $\mathcal{V}$ and $\mathcal{E}$ are sets containing nodes and edges, respectively. A node $i \in \mathcal{V}$ represents an entity, and an edge $(i, j) \in \mathcal{E}$ defines a directed relation from node $i$ to node $j$. Denote the neighboring set of node $i$ as $\mathcal{N}_{i}=\{j \in \mathcal{V} \mid(j, i) \in$ $\mathcal{E}\}$. The attributes of node $i$ and edge $(i, j)$ are characterized by $\mathbf{v}_{i}$ and $\mathbf{e}_{i j}$, respectively.  
 \begin{figure}[ht]
\setlength{\abovecaptionskip}{0pt}
\setlength{\belowcaptionskip}{0pt}
\centering
\includegraphics[width= 0.4\textwidth]{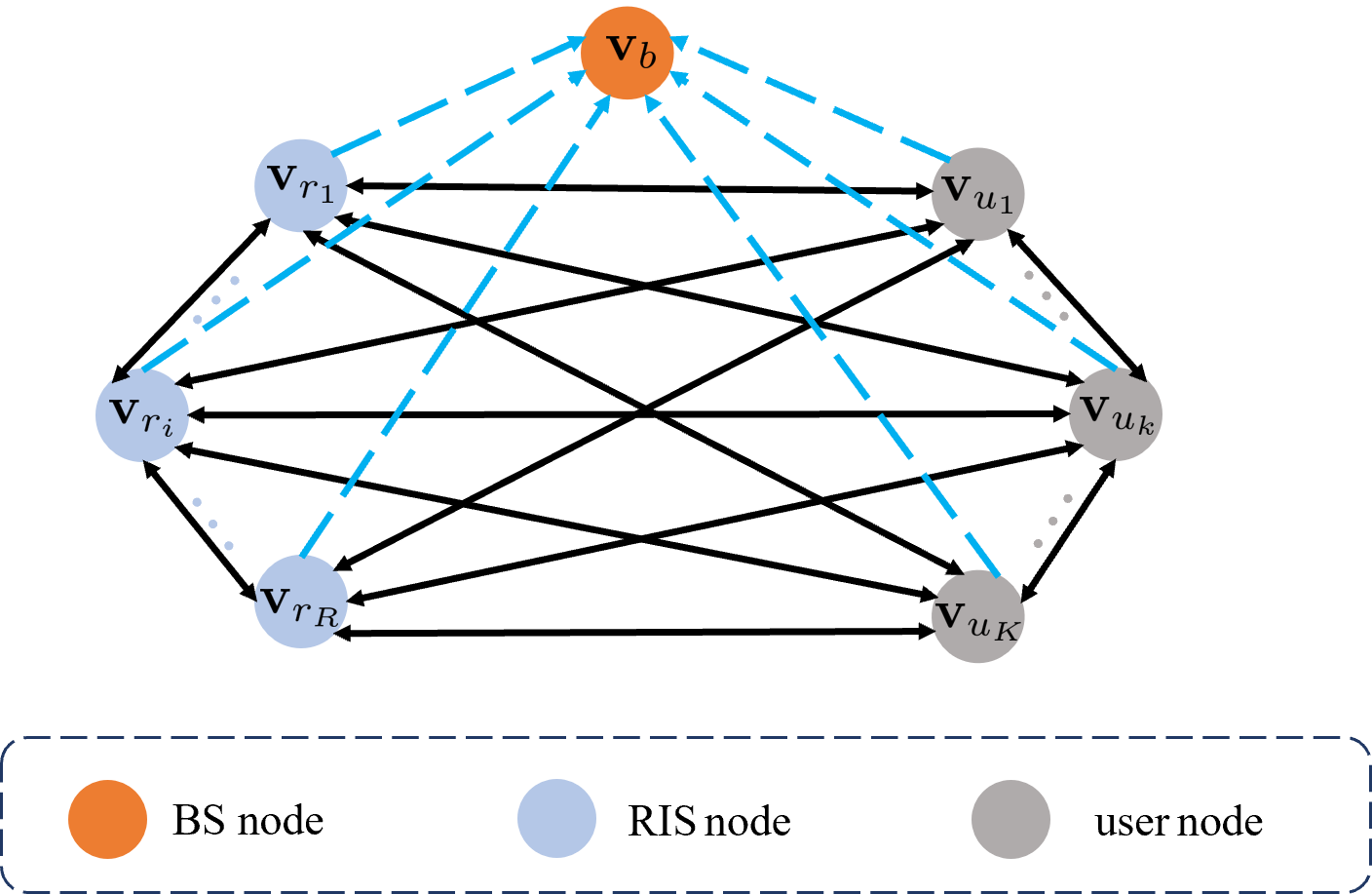}
\DeclareGraphicsExtensions.
\caption{{  The graph representation in the considered heterogeneous GNN, where the black solid edge indicates that information can be exchanged bidirectionally, and the blue dotted line indicates the directional output to the BS node.}}
\label{2}
\end{figure}

\begin{figure*}[hbt]
\centering
\includegraphics[scale=.54]{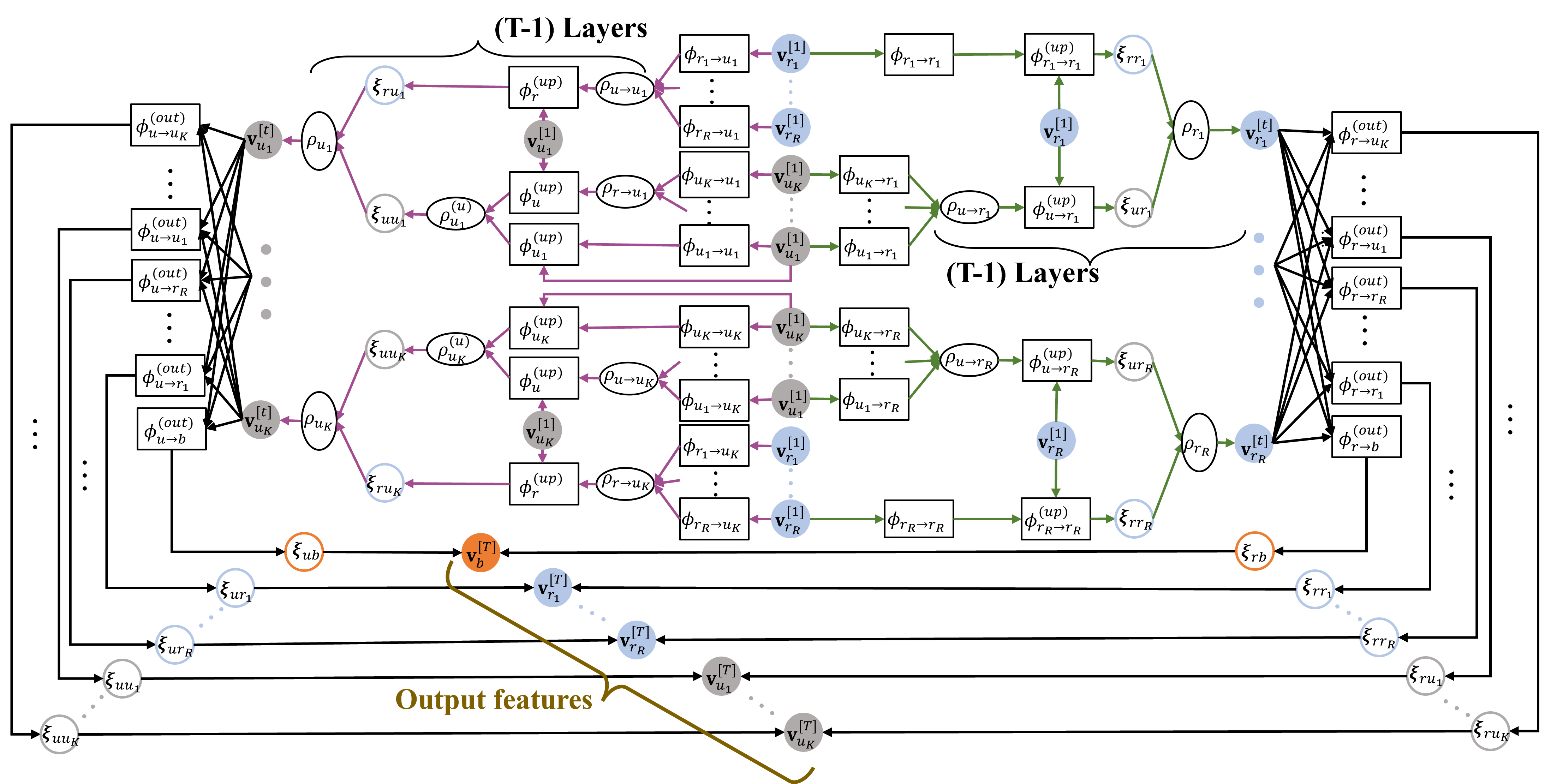}
\caption{The message-passing paradigm in the heterogeneous GNN network. The purple line on the left represents the update process for the features of the user nodes and the green line on the right represents the features for the update process of the RIS nodes. In addition, the black line at the bottom denotes the output layers for the features of the final nodes.}
\label{3}
\end{figure*}
{ 
Specifically, the optimization problem P1 encompasses a variety of variables each bearing distinct characteristics, including $\mathbf{F}$ related to the beamforming design of the BS, ${\mathbf{\Phi}}_{i}$ related to the phase configuration of the RIS$_i$, and $\mathbf{U}$ related to the association scheme of the users. To effectively manage this diversity,  we introduce a heterogeneous GNN model \cite{zhang2019heterogeneous},  which is adept at encapsulating a wealth of information through structural relations among many types of nodes and the unstructured content inherent to each node, as illustrated in Fig. \ref{2}.}
 A heterogeneous graph $\mathcal{G}=$ $(\mathcal{V}, \mathcal{E}, \mathcal{O}_\mathcal{V}, \mathcal{R}_\mathcal{E})$, it is characterized by two additional sets, which are the predefined type of nodes $\mathcal{O}_\mathcal{V}$ and type of nodes $\mathcal{R}_\mathcal{E}$ with $|\mathcal{O}_\mathcal{V}|+|\mathcal{R}_\mathcal{E}|>2$ \cite{schlichtkrull2018modeling}.  
 The set $\mathcal{O}_\mathcal{V}$ includes the {BS node} $\mathbf{v}_{b}$, {RIS$_i$ node} $\mathbf{v}_{r_i}$ and {user$_k$ node} $\mathbf{v}_{u_k}$.
 
 { 
 According to the considered wireless communication scenario, the attributes of { user$_k$ node } include the weight $w_k$, noise variance $\sigma^2$ as well as the RIS association matrix $\mathbf{u}_k$. Similarly, the attributes {  of the BS node and the RIS$_r$ node} include the beamforming matrix $\mathbf{F}$ and the phase configurations ${\mathbf{\Phi}}_{i}$, respectively.
 {  In addition, the edges in the network
are only used to represent the connection between two nodes without attributes. 
 }  } Therefore,  the heterogeneous GNN utilizes only one type of edge, leading to the omission of the types of edges set $\mathcal{R}_\mathcal{E}$ in this study.
 Given the heterogeneous GNN $\mathcal{G}=$ $(\mathcal{V}, \mathcal{E}, \mathcal{O}_\mathcal{V})$, the objective of this paper is to find a policy  $p_\Theta(\cdot)$ with parameters $\Theta$, mapping the graph $\mathcal{G}$ to the optimal ${\mathbf{F},\mathbf{\Phi}_i, \mathbf{U}}$ in multi-RIS aided communication systems. 

 \subsection{The Initialization Layer of the Heterogeneous GNN}

 The initial layer of the heterogeneous GNN takes the CSI  data as the input features for the whole heterogeneous GNN. In order to better learn the phase configuration of each RIS $\Phi_i$, we extract the phase shift matrix $\boldsymbol{\theta}_i$ and the equivalent channel $\mathbf{h}_k$ can be expressed as  
 \begin{align}\label{CSI_input}
    \mathbf{h}_{k} 
   &= \sum_{i \in \mathcal{R}} u_{k, i} {\boldsymbol{\theta}}_i
   \mathrm{diag}(\mathbf{h}_{ik}) \mathbf{G}_{i} \\ \nonumber
   &= \sum_{i \in \mathcal{R}} u_{k, i} {\boldsymbol{\theta}}_i \mathbf{H}_{ik}, 
 \end{align}
where $\mathbf{H}_{ik} \in \mathbb{C}^{M \times N_t}$ denotes the cascaded channel and ${\boldsymbol{\theta}}_i = \left[\theta_{i,1},\cdots,\theta_{i,M}\right] \in \mathbb{C}^{1 \times M}$ represents the phase coefficient of RIS$_i$. 
{  Thus, the CSI $\mathbf{H}_{ik}$ from all the RISs for each user $k$ is concatenated as $\overline{\mathbf{H}}_{ik}\in \mathbb{C}^{RM \times N_t}$, which is considered as the input feature $\mathbf{v}_{u_k}^{[0]}$ for each user node, i.e., $\mathbf{v}_{u_k}^{[0]} = \left[ \operatorname{Re}(\overline{\mathbf{H}}_{ik}), \operatorname{Im}(\overline{\mathbf{H}}_{ik}) \right]$.
}
Given the input feature representation $\mathbf{v}_{u_k}^{[0]}$, we use an encoder block $f_\Theta^{[1]}(\cdot)$, consisting of one layer of fully connected neural networks and a Leaky ReLU activation function \cite{maas2013rectifier}, to produce $\mathbf{v}_{u_k}^{[\mathrm{1}]}$ for the user$_k$ node and $\mathbf{v}_{r_i}^{[\mathrm{1}]}$ for the RIS$_i$ node. 

In a multi-RIS multi-user communication network, each user experiences interference from the other users. In order to analyze the interference, the channels of the interfering links are aggregated to update the features through the encoder block $f_{\Theta_u}^{[1]}(\cdot)$, which is given as 
\begin{align}\label{i1}
\mathbf{v}_{u_k}^{[\mathrm{1}]}  &=f_{\Theta_u}^{[1]}\left([\mathbf{v}_{u_1}^{[\mathrm{0}]}, \cdots, \mathbf{v}_{u_k'}^{[\mathrm{0}]}, \cdots,  \mathbf{v}_{u_K}^{[\mathrm{0}]}]\right), \quad k' \in  {\mathcal{K}}, k' \neq k. 
\end{align}

For each RIS node, all the users receive the signals reflected through the RIS, so all the channels of the users contain an equal amount of information about the RIS. The update for the RIS node by the encoder block $f_{\Theta_r}^{[1]}(\cdot)$ is defined as 
\begin{align}\label{i2}
\mathbf{v}_{r_i}^{[\mathrm{1}]} =f_{\Theta_r}^{[1]}\left([\mathbf{v}_{u_1}^{[\mathrm{0}]}, \cdots, \mathbf{v}_{u_k}^{[\mathrm{0}]}, \cdots,  \mathbf{v}_{u_K}^{[\mathrm{0}]} ]\right), \quad k \in \mathcal{K}.
\end{align}

\subsection{Message-Passing Paradigm}

The graph network (GN) block is adopted as the basic computation unit over graphs \cite{battaglia2018relational}, which includes the update functions $\phi$ and aggregation functions $\rho$, as illustrated in Fig. \ref{3}. The message-passing for the RIS nodes and user nodes are marked in green lines and purple lines, respectively. The update of the representation vector in the (T-1)-th layer is based on combining its previous representation and the aggregation of the representations from its neighboring nodes.
 
\subsubsection{Update for RIS Nodes}

The updated feature $\mathbf{v}_{r_i}^{[t]}$ for the RIS$_i$ node at step $t$ ($2 \leq t \leq T-1$)  is obtained by combining the features from the last state $\mathbf{v}_{r_i}^{[t-1]}$ and all the user nodes  $\mathbf{v}_{u_k}^{[t-1]}, \forall k \in \mathcal{K}$. Then, the message $\boldsymbol{\xi}_{rr_i}$ from its own last state and the message $ \boldsymbol{\xi}_{ur_i} $ from the user nodes are given as 
\begin{align}\label{m1}
\boldsymbol{\xi}_{rr_i} =  \phi_{r_i \rightarrow r_i}^{({\rm{up}})}(  \boldsymbol{\xi}_{r_i}, \mathbf{v}_{r_i}^{[t-1]}),
\end{align}
\begin{align}
 \boldsymbol{\xi}_{ur_i} = \phi_{u \rightarrow r_i}^{({\rm{up}})}( \rho_{u \rightarrow r_i}( \boldsymbol{\xi}_{u_1} ,\cdots, \boldsymbol{\xi}_{u_K}), \mathbf{v}_{r_i}^{[t-1]}),
\end{align}
with
\begin{align}
 \boldsymbol{\xi}_{\rm{src}} = \phi_{{\rm{src}} \rightarrow r_i}(\mathbf{v}_{\rm{src}}^{[t-1]}), {\rm{src}} \in \{r_i, u_1, \cdots, u_K\}, 
\end{align}
{ where \(\phi_{\mathrm{src}\to r_i}(\cdot)\) and \(\phi_{\mathrm{src}\to r_i}^{(\mathrm{up})}(\cdot)\) are two-layer multi-layer perceptrons (MLPs) for message combining and updating, respectively, with the superscript \((\mathrm{up})\) denoting “update”.}
 Besides, $\rho_{u \rightarrow r_i} $ is the mean aggregation function for all the different user nodes \cite{fey2019fast}, i.e., $\rho_{u \rightarrow r_i}( \boldsymbol{\xi}_{u_1},\cdots, \boldsymbol{\xi}_{u_K}  ) = {1}/{K} \sum_{ k = 1}^K  \boldsymbol{\xi}_{u_k}$. 
 
 Then, in the considered heterogeneous GNN, the updated RIS node $\mathbf{v}_{r_i}^{[t]}$ aggregates the feature information from different-type neighbors, which is given as 
\begin{align}
  \mathbf{v}_{r_i}^{[t]} = \rho_{r_i}( \boldsymbol{\xi}_{rr_i},  \boldsymbol{\xi}_{ur_i}) + \mathbf{v}_{r_i}^{[t-1]},  
\end{align}
where $\rho_{r_i} (\cdot)$ denotes the operation of mean aggregation, which aggregates the features from the RIS node and user nodes. Besides, a residual operation is added in this step to avoid the degradation and gradient explosion of the considered network \cite{2016Deep}. Thus, the node features $\mathbf{v}_{r_i}^{[t-1]}$ at the last iteration can be retained through this residual operation.

 \subsubsection{Update for User Nodes}
 
The updated feature $\mathbf{v}_{u_k}^{[t]}$ is obtained by combining the features from all the user nodes and all the RIS nodes. In the same manner, the message $\boldsymbol{\xi}_{ru_k}$ combined from the RIS nodes and the message $ \boldsymbol{\xi}_{uu_k} $ combined from the user nodes are given as 
\begin{align}
 {\boldsymbol{\xi}}_{ru_k} &= \phi_{r \rightarrow u_k}^{({\rm{up}})}( \rho_{r \rightarrow u_k}( \boldsymbol{\xi}_{r_1} ,\cdots, \boldsymbol{\xi}_{r_R}), \mathbf{v}_{u_k}^{[t-1]}),
\end{align}
\begin{align}
 {\boldsymbol{\xi}}_{uu_k} = \rho_{u_k}^{(u)} ( \phi_{u \rightarrow u_k}^{({\rm{up}})} (\boldsymbol{\xi}_{u}),  \phi_{u_k\rightarrow u_k}^{(up)}(\boldsymbol{\xi}_{u_k},
 \mathbf{v}_{u_k}^{[t-1]})) ),
\end{align}
where both $ \phi_{{\rm{src}} \rightarrow u_k}(\cdot)$ and $\phi_{ {\rm{src}}\rightarrow u_k}^{({\rm{up}})}(\cdot)$ are two-layer MLPs for the message combination and update process of $\mathbf{v}_{u_k}^{[t]}$, respectively. Moreover, $\boldsymbol{\xi}_{u} = \rho_{u \rightarrow u_k}( \boldsymbol{\xi}_{u_1},\cdots, \boldsymbol{\xi}_{u_K})$ denotes the message aggregated from all the user nodes. The aggregation function $\rho_{u \rightarrow u_k}(\cdot)$ is a max aggregation function, i.e.,  $\rho_{u \rightarrow u_k}( \boldsymbol{\xi}_{u_1},\cdots, \boldsymbol{\xi}_{u_K}) = \max (\boldsymbol{\xi}_{u_1},\cdots, \boldsymbol{\xi}_{u_K})$, {\color {black} where the max-pooling is adopted as it performs well and effectively captures the strongest interference signal among the users \cite{jiang2021learning}. 
} Besides, $\boldsymbol{\xi}_{\rm{src}} = \phi_{{\rm{src}} \rightarrow r_i}(\mathbf{v}_{\rm{src}}^{[t-1]}), {\rm{src}} \in \{r_1, \cdots, r_R, u_1, \cdots, u_K\}$ denotes the message from each neighbor node and $\rho_{r \rightarrow u_k}$ is the mean aggregation function. 

Then, in heterogeneous GNNs, the updated user node $\mathbf{v}_{u_k}^{[t]}$ aggregates the feature information from different neighbors, which is given as
\begin{align}\label{m2}
  \mathbf{v}_{u_k}^{[t]} = \rho_{u_k}( \boldsymbol{\xi}_{ru_k},  \boldsymbol{\xi}_{uu_k}) + {\mathbf{v}}_{u_k}^{[t-1]},  
\end{align}
where $\rho_{u_k}(\cdot)$ denotes the operation of mean aggregation function.
 
\subsubsection{Output Layers for Node Features}

 To obtain the final feature of each node $\mathbf{v}_{b}^{[T]}$, $\mathbf{v}_{r_i}^{[T]}$ and $\mathbf{v}_{u_k}^{[T]}$, a decoder layer is used to output all the optimized features, which is given as
\begin{align} \label{o3}
  \mathbf{v}_{\rm{dst}}^{[T]} &= \rho_{ \rm{dst}}^{\rm{(out)}}( \boldsymbol{\xi}_{u{\rm{dst}}}, \boldsymbol{\xi}_{r{\rm{dst}}}), \nonumber\\
  {\rm{dst}} &\in \{b, r_1, \cdots, r_R, u_1, \cdots, u_K\}, 
\end{align}
where $\boldsymbol{\xi}_{u{\rm{dst}}} = \phi_{\rm{u} \rightarrow \rm{dst}}^{(\rm{out})} (\mathbf{v}_{u_1}^{[T-1]}, \cdots, \mathbf{v}_{u_K}^{[T-1]})$ and $\boldsymbol{\xi}_{r{\rm{dst}}} = \phi_{\rm{r} \rightarrow \rm{dst}}^{(\rm{out})} (\mathbf{v}_{r_1}^{[T-1]}, \cdots, \mathbf{v}_{r_R}^{[T-1]})$ denotes the destination node features extracted from the user nodes and RIS nodes, respectively.  As shown by the black line in Fig. \ref{3},  the final state $\mathbf{v}_{\rm{{{dst}}}}^{[T]}$ is obtained by a mean aggregation function $\rho_{\rm{dst}}^{(\rm{out})}(\cdot)$. {
According to the defined types of nodes already mentioned, the BS node outputs the beamforming matrix $\overline{\mathbf{F}}$ as $\mathbf{v}_{\rm{b}}^{[T]} = [ \operatorname{Re}\{\overline{\mathbf{F}}\}, \operatorname{Im}\{\overline{\mathbf{F}}\}]$. Similarly, the RIS$_r$ node and the user$_k$ node output the phase configuration matrix $\overline{\mathbf{\Phi}}_i$ and the association scheme $\overline{\mathbf{u}}_k$, respectively. Specifically, $\mathbf{v}_{{\rm{r}}_i}^{[T]} = [ \operatorname{Re}\{\mathbf{\Phi}_i\}, \operatorname{Im}\{\mathbf{\Phi}_i\}]$ and $\mathbf{v}_{u_k}^{[T]} = \overline{\mathbf{u}}_k$.
}
{ 
After obtaining the final state of each node, the node features are fed into the normalization layers to meet the constraints in P1 (12b) $\sim$ (12e). Then, the WSR-related loss functions are used to update the network parameters to obtain the desired variables in P1.}

Suppose that $T$ iterations are executed during the training process, and we define each update as HGN$_t, \forall t \in T$. {  An encoder-process-decoder architecture is utilized to facilitate the training \cite{battaglia2018relational}, including one encoder block HGN$_1$ to initialize the node features through (\ref{i1}) and  (\ref{i2}), one decoder block HGN$_T$ to output the node features through (\ref{o3}), and one core block (a concatenation of blocks from HGN$_2,\cdots$, HGN$_t,\cdots$, HGN$_{T-1}$) to implement the message-passing paradigm through (\ref{m1}) to (\ref{m2}).} After that, we can obtain the final result for the optimal $\mathbf{F}$, $\mathbf{U}$ and $\mathbf{\Phi}_i, \forall i \in \mathcal{R}$.

\subsection{Model Architecture of the Heterogeneous GNN}

In addition to the message-passing paradigm, the proposed heterogeneous GNN needs some additional layers to complete the whole learning process, and the functions of these layers are listed as follows.

\subsubsection{The Normalization Layers}

Typical neural networks operate on real numbers, hence it is not possible to learn the mapping of the channel information to the values of { ${\mathbf{F}} $ and ${\mathbf{\Phi}_i}$}, which are all complex values in wireless communication systems. Instead, we learn the vector consisting of the real and imaginary parts of the beamforming matrix ${\mathbf{F}}$ and the phase configuration matrices $\mathbf{\Phi}_i, \forall i \in \mathcal{R}$. Therefore, before calculating the WSR loss function, additional layers are necessary to merge the real-valued results into complex-valued results. 

\begin{itemize}
\item To satisfy the constraint on the transmit power in (12b), a normalization layer is employed at the output feature {  of the BS node $\mathbf{v}_{\rm{b}}^{[T]} = [ \operatorname{Re}\{\overline{\mathbf{F}}\}, \operatorname{Im}\{\overline{\mathbf{F}}\}]$}, and the function of this layer can be expressed as
\begin{align}
{\mathbf{F}}=\sqrt{{P}_{{\max}}} \frac{\operatorname{Re}\{\overline{\mathbf{F}}\}+j  \operatorname{Im}\{\overline{\mathbf{F}}\}}{\sqrt{\left\|\operatorname{Re}\{\overline{\mathbf{F}}\}\right\|_{\mathcal{F}}^2+\left\|\operatorname{Im}\{\overline{\mathbf{F}}\}\right\|_{\mathcal{F}}^2}}.   
\end{align}
 
\item Similarly, the phase configuration of each RIS needs to fulfill the unit-modulus constraint (12c), and a normalization layer is utilized to tackle the real and imaginary components of each reflection coefficient ${\overline{\theta}}_{i,m}$ {  in the output feature of the RIS$_i$ node 
$\mathbf{v}_{{\rm{r}}_i}^{[T]} = [ \operatorname{Re}\{\overline{\mathbf{\Phi}}_i\}, \operatorname{Im}\{\overline{\mathbf{\Phi}}_i\}], \forall i \in \mathcal{R}$ }. Thus, the function can be represented as
\begin{align}
{{\theta}}_{i,m} = \frac{\operatorname{Re} \{ {\overline{\theta}}_{i,m} \}+ j  \operatorname{Im} \{ {\overline{\theta}}_{i,m} \} }{ \sqrt{ |\operatorname{Re} \{  {\overline{\theta}}_{i,m} \}|^2 + |\operatorname{Im} \{  {\overline{\theta}}_{i,m} \} |^2 } }.
\end{align} 
\end{itemize}

\subsubsection{The RIS  Association Scheme}

Each user chooses a suitable RIS as the reflection link to enhance the link quality, as dictated by constraint (12d). Thus, we utilize a SoftMax layer on the output feature {  of each user node $\mathbf{v}_{u_k}^{[T]} = \overline{\mathbf{u}}_k$} to implement the association scheme. {  For each element $\overline{{u}}_{k,i}$ in  $\overline{\mathbf{u}}_k$, the SoftMax score can be calculated by}
\begin{align}
    c_{u_{k,i}} = \operatorname{SoftMax}\left(\overline{{u}}_{k,i}\right)=\frac{e^{\overline{{u}}_{k,i}}}{\sum_{i=1}^R e^{\overline{{u}}_{k,i}}}.
\end{align}

After obtaining the SoftMax score $c_{u_{k,i}}$, we sort the scores of each user, i.e., each column in $\mathbf{U}$. 
{  Leveraging the Softmax function for the RIS association scheme inherently allows for the possibility of associating multiple RISs with each user. We select the positions of the { largest} values and set them to 1, indicating that the user$_k$ selects {  the most appropriate RIS to assist the communication links.  Therefore, the selection scheme is obtained }}
\begin{equation}
   {u}_{k,i}=\left\{\begin{array}{l}
1, \quad \text{if} \quad c_{u_{k,i}} = \max\limits_{{ i \in \mathcal{R}}}(c_{u_{k,i}}), \forall k \in \mathcal{K} \\
0, \quad \text {otherwise.}
\end{array}\right.
\end{equation}

\begin{figure}[ht]
\setlength{\abovecaptionskip}{0pt}
\setlength{\belowcaptionskip}{0pt}
\centering
\includegraphics[width= 0.4\textwidth]{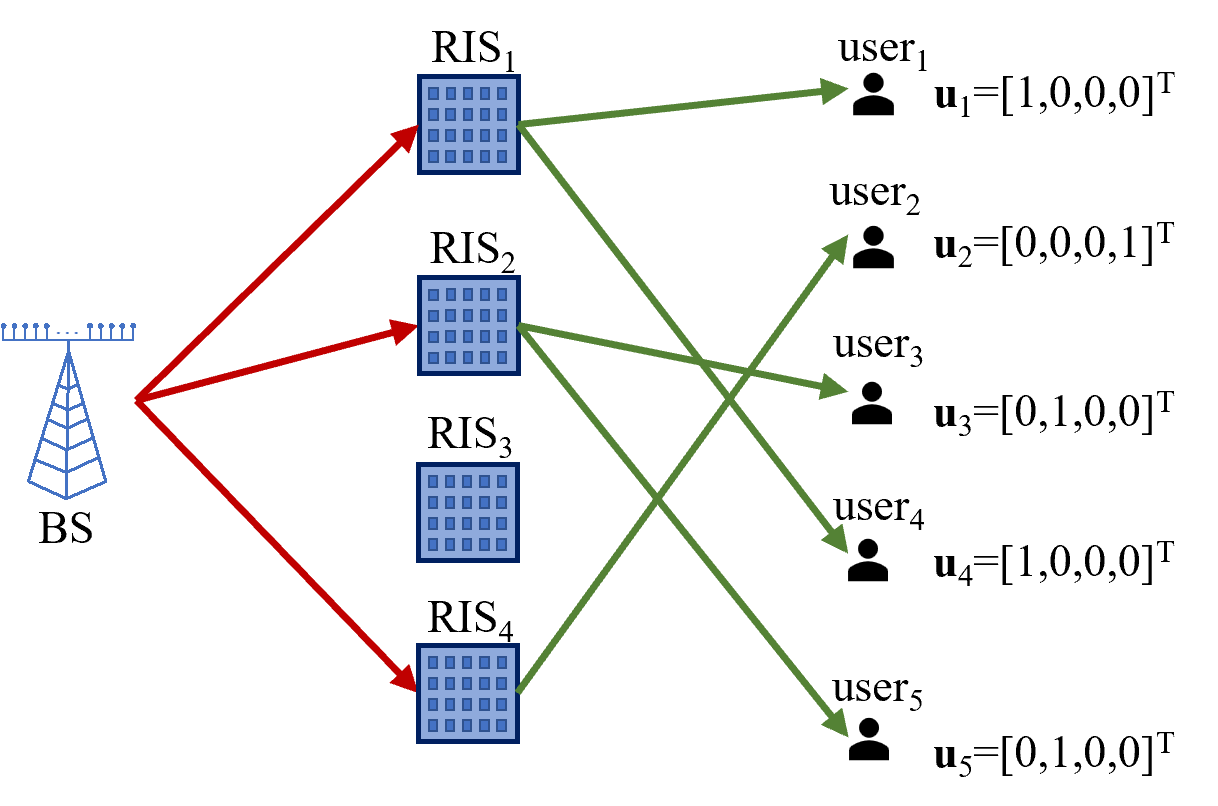}
\DeclareGraphicsExtensions.
\caption{{  An example of the RIS association scheme in the multi-RIS multi-user system, considering 4 RISs and 5 users.}}
\label{4}
\end{figure}
{ 
For ease of understanding, a network deployment with 4 RISs and 5 users is considered as an example to explain the structure of the matrix $\mathbf{U}$ in Fig. \ref{4}. 
It is apparent that each user selects only one RIS as the cascaded communication link, whereas each RIS can serve multiple users or none of them. For instance, $\mathbf{u}_1 = [1,0,0,0]^T$ means that user$_1$ chooses RIS$_1$ as the cascaded link. Meanwhile, we notice that the third row of each user node $\mathbf{u}_k,\forall k \in \mathcal{K}$ is 0, which means that no user chooses RIS$_3$ as a cascaded link. Accordingly, RIS$_3$ is set to the ‘OFF’ state, which implies that it does not reflect any signals to any users.}

\begin{figure*}[t]
\centering
\includegraphics[scale=.38]{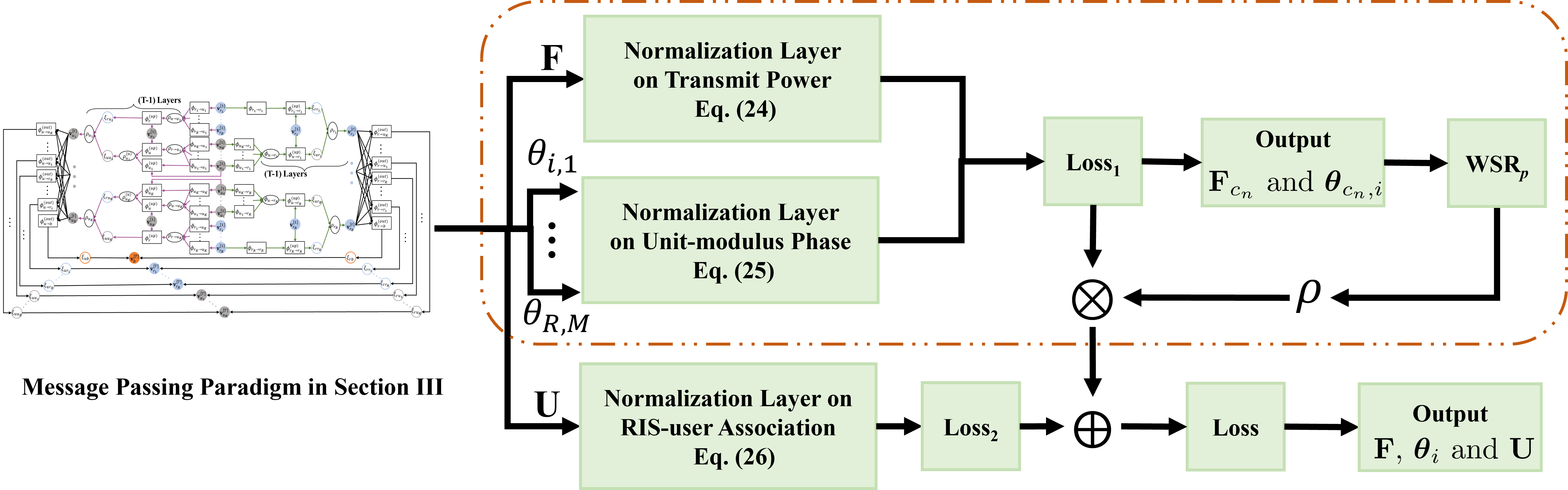}
\caption{The overall structure of the proposed heterogeneous GNN. The left part is the message-passing paradigm, which has been illustrated in Section III. The orange rectangle identifies the pre-training process for the maximization problem with the predetermined RIS association scheme.}
\label{structure}
\end{figure*}

\section{{  The Training Process of the Heterogeneous GNN in Multi-RIS Multi-user Systems}}

{\color {black} Based on the message-passing paradigm and the normalization layers introduced in the previous section, we focus on maximizing the WSR in multi-RIS aided communication systems.  
{  
 Aiming to address the challenge of learning continuous and discrete variables in the maximization problem, a pre-training process is adopted to obtain the penalty term. Therefore,  the overall training processes and the pre-training process are elaborated in the following section.
} }

\subsection{{ The Training Process of the Heterogeneous GNN}}
{  As defined in problem P1, }
in addition to the beamforming matrix and phase configuration at the RISs, the heterogeneous GNN is required to output the RIS association scheme $\mathbf{U}$.
However, the association matrix is composed of binary elements, which are discrete variables. Thus, the process to determine the optimal value of $\mathbf{U}$, $\mathbf{F}$, and $\mathbf{\Phi}_i, \forall i \in \mathcal{R}$ is very difficult to learn. In order to better learn the RIS association matrix,  it is important to have a suitable optimization criterion.
{ 
Therefore, two different loss functions are adopted to solving the maximization problem P1. One is related to the WSR of the multi-RIS aided communication systems for maximizing the target of the considered problem, which is defined as 
\begin{align}
\label{loss1}
{\operatorname{Loss}}_1 
&= -\sum_{k = 1}^{K} w_{k}{\rm{log}}_2(1 +  {\mathrm{SINR}}_{ k})  \\  
&= -\sum_{k = 1}^{K} w_{k}{\rm{log}}_2(1+ \frac{\left| {\mathbf{h}}_{{k}} \mathbf{f}_{{k}}\right|^{2}}{ \sum\limits_{ \it{j}\in \mathcal{K}, \it{j} \neq {k}}^{{K}} \left|
    \mathbf{h}_{{k}} \mathbf{f}_{{j}}
    \right|^{2}+\sigma^{2}}).\nonumber
\end{align}}
\begin{algorithm}[htb]
\caption{ The overall process of the heterogeneous GNN}
\label{alg:Framwork}
\begin{algorithmic}[1] 
\REQUIRE ~~\\ 
    The heterogeneous graph $\mathcal{G} = (\mathcal{V}, \mathcal{E}, \mathcal{O}_{\mathcal{V}} )$;\\
    The training and testing samples $ \mathcal{S} = \left\{\mathbf{H}_{ik}\right\}, \forall i \in \mathcal{R}, k \in \mathcal{K}$. 
\ENSURE ~~\\ 
    The final node embedding feature $\mathbf{F}$, $\mathbf{\Phi}_i, \forall i \in \mathcal{R}$, and $\mathbf{U}$; \\ 
    \FOR {epoch $  = 1, \cdots, N_e $}
    \STATE Initialize the node features via the encoder block HGN$_0$  from samples for all the RIS nodes and user nodes using (14) and (15);
    \FOR {$t =  1, 2, \cdots, T-1 $}
        \FOR { all the user nodes and RIS nodes}
        \STATE Combine the message from the different source node features for all the RIS nodes and user nodes using (16)-(18) and (20)-(21);
           \STATE Aggregate the message from different-type neighbors for all the RIS nodes and user nodes using (19) and (22);
        \ENDFOR
    \STATE Update the node feature $\mathbf{v}_{p}^{[t+1]}, 
 \forall p \in {\mathcal{O}}_{\mathcal{V}}$ using (23);
    \ENDFOR
     \STATE Calculate the loss function using (\ref{loss}):\\ $\operatorname{Loss}  = - \sum_{k = 1}^{K} \left( w_{k}\rm{log}_2(1+ {\mathrm{SINR}}_{\it{k}})+\eta  \mathbf{u}_{k}^{(l)}  \log { \mathbf{c}_{u_{k}}}\right)$;
     \STATE Perform backpropagation and update the parameters of the heterogeneous GNN;
    \ENDFOR
\RETURN $\mathbf{F}, \mathbf{\Phi}_i, \forall i \in \mathcal{R}$, and $\mathbf{U}$. 
\end{algorithmic}
\end{algorithm}
{\color {black}Then, to achieve more efficient matching based on the nearest-distance principle, we employ the CE loss function to guide the learning process, defined as}
\begin{align}
 \operatorname{Loss}_2 (\mathbf{u}_k^{(\mathrm{l})}, { \mathbf{c}_{u_{k}}})&=  { -} \sum_{k=1}^K \mathbf{u}_{k}^{(\mathrm{l})}  \log { \mathbf{c}_{u_{k}}},
 \label{loss2}
\end{align}
where ${\mathbf{u}}_{k}^{(\mathrm{l})}$ and ${ \mathbf{c}_{u_{k}}}$ denote the distance-dependent label and the output for the user$_k$, respectively. {  
The process of obtaining the distance-dependent label ${\mathbf{u}}_{k}^{(\mathrm{l})}$ requires the calculation of the distances between the users and RISs using their location information from the dataset. For each user $k$, we identify the nearest RIS, denoted by the index $\hat{r}$. In the label vector $\mathbf{u}_k^{(\mathrm{l})}$, we set the element corresponding to $\mathbf{u}_{k,\hat{r}}$ to $1$, which indicates the association with the nearest RIS, while all the other elements are set to $0$, which implies no association.
}

Based on the output variables and the CSI information, the loss function is defined as a weighted sum of Loss$_1$ and Loss$_2$, which can be expressed as
\begin{align}
 \operatorname{Loss} &=  \operatorname{Loss}_1 + \eta \operatorname{Loss}_2  \nonumber \\
&= -\sum_{k = 1}^{K} \left( w_{k}\rm{log}_2(1+ {\mathrm{SINR}}_{\it{k}}) + \eta\mathbf{u}_{k}^{(l)}  \log { \mathbf{c}_{u_{k}}}\right).
 \label{loss}
\end{align}

{  The first term in the loss function, \(\operatorname{Loss}_1\), optimizes the overall system’s weighted sum rate, while the second term, \(\operatorname{Loss}_2\), is a distance-dependent loss that encourages the selection of the nearest RIS.} The coefficient $\eta$ is a penalty term, and its role is to strike a balance between the two components of the loss function, which can be calculated by
\begin{align}
    \eta = \frac{{\rm{WSR}}_p}{{\rm{WSR}}_{p_0}},
\end{align}
where ${{\rm{WSR}}_{p}}$ is the result in the pre-training process by using the loss function in (\ref{loss1}) and ${{\rm{WSR}}_{p_0}}$ is the result under the distance-dependent label $\mathbf{u}_{k}^{(l)}$ when $P_{ {\max}} = 30 $ dBm. {  For ease of understanding, the overall process of the heterogeneous GNN for solving   P1 is shown as Algorithm 1 and the pre-training process is given in the next subsection.}

\subsection{  The Pre-training Process for the Penalty Term}
\begin{algorithm}[htb]
\caption{ {  The pre-training process}}
\label{alg:Framwork1}
\begin{algorithmic}[1] 
\REQUIRE ~~\\ 
    { A predetermined scheme ${\mathbf{U}}^{({\rm l})} = [\mathbf{u}_{1}^{(\rm l)}, \cdots, \mathbf{u}_{K}^{(\rm l)}]$};\\
    The heterogeneous graph $\mathcal{G} = (\mathcal{V}, \mathcal{E}, \mathcal{O}_{\mathcal{V}} )$;\\
    The training and testing samples $ \mathcal{S} = \left\{\mathbf{H}_{ik}\right\}, \forall i \in \mathcal{R}, k \in \mathcal{K}$. \\  
\ENSURE ~~\\ 
    The final node embedding feature ${\mathbf{F}}^{({\rm l})}$ and ${\boldsymbol{\theta}}_{i}^{({\rm l})}, \forall i \in \mathcal{R}$. \\ 
    \FOR {epoch $  = 1, \cdots, N_e $}
        \STATE Initialize the node features via the encoder block HGN$_0$ as $\mathbf{v}_{\rm{src}}^{[1]}$ from the training sample set for all the RIS nodes and user nodes using (14) and (15);
        \FOR {$t = 2, \cdots, T-1 $}
            \FOR { all the user nodes and RIS nodes}
                \STATE Combine the message from the different source node features for all the RIS nodes and user nodes using (16)-(18) and (20)-(21);
               \STATE Aggregate the message from different-type neighbors for all the RIS nodes and user nodes using (19) and (22);
            \ENDFOR
        \STATE Update the features of the BS node $\mathbf{v}_{b}^{[T]}$ and RIS nodes $\mathbf{v}_{r_i}^{[T]}, \forall i \in \mathcal{R}$ using (23);
    \ENDFOR
     \STATE Calculate the WSR loss using (\ref{loss1}):\\
     $\operatorname{Loss}  = -  \sum_{k = 1}^{K} w_{k} {\rm{log}}_2(1+ {\mathrm{SINR}}_{ k}^{\rm(n)})  $;
     \STATE Perform backpropagation and update parameters in the heterogeneous GNN;
     \ENDFOR
\RETURN  ${\mathbf{F}}^{\rm(l)}$ and ${\boldsymbol{\theta}}_{i}^{\rm(l)}, \forall i \in \mathcal{R}$. 
\end{algorithmic}
\end{algorithm}
{ {

To obtain ${{\rm{WSR}}{p}}$, we introduce a pre-training process where the learning is based on the distance-dependent label $\mathbf{u}_{k}^{({\rm l})}$, assuming all users connect to their nearest RIS already. Thus, the message-passing paradigm remains the same as illustrated in Section III-C, but without returning $\mathbf{U}$, as the orange rectangle in Fig. \ref{structure}.
Based on the predetermined $\mathbf{u}_{k}^{({\rm l})}$, the cascaded channel $ \mathbf{h}_{k}^{(\rm l)}$  can be rewritten as
\begin{align}
 \mathbf{h}_{k}^{\rm(l)}   =  \sum_{i \in \mathcal{R}} u_{k,i}^{({\rm l})} {\boldsymbol{\theta}}_{i}^{\rm(l)} \mathbf{H}_{ik}, 
 \label{h_n}
\end{align}
where $\mathbf{H}_{ik} = \mathrm{diag}(\mathbf{h}_{ik}) \mathbf{G}_{i}$ is the input of the heterogeneous GNN. 
Therefore, the WSR maximization problem, given the predetermined association scheme $\mathbf{u}_{k}^{({\rm l})}$, aims to find the optimal $\mathbf{F}^{{({\rm l})}}$ and $\boldsymbol{\theta}_i^{{({\rm l})}}$, which can be expressed as 
\begin{subequations}
\begin{align}
\text { P2: }
{\mathop{\max}\limits_{\mathbf{F}^{{({\rm l})}},  \boldsymbol{\theta}^{{({\rm l})}}_{i}, \forall i \in \mathcal{R} }} & \sum_{k = 1}^{K}w_k {\rm{log}}_2(1+ {\rm{SINR}}_k^{({\rm l} )}) \\
\text { s.t. } 
& \left\|\mathbf{F}^{{({\rm l})}} \right\|_\mathcal{F}^{2} \leq P_{max} \\
& \left|\boldsymbol{\theta}_{i,m}^{({\rm l})}\right| = 1,  \forall m = 1,2, \cdots, M. 
\label{P3}
\end{align}
\end{subequations}

For ease of understanding, the overall pre-training process of the heterogeneous GNN for solving the maximization problem P2 is shown as Algorithm 2, i.e. the training process in the orange rectangle in Fig. \ref{structure}. 
}}

\section{Numerical Results}

\subsection{Simulation Setup}
\begin{table}[h]
    \caption{  SIMULATION AND TRAINING HYPERPARAMETERS}
    \label{tab:hyperparams}
    \centering
    \scalebox{0.82}{
        \begin{tabular}{llll}
            \toprule
            \textbf{ Parameter} & \textbf{ Value} & \textbf{ Parameter} & \textbf{ Value} \\
            \midrule
             Reflecting elements ($M$) &  16 &  Transmit antenna elements ($N_t$) &  8 \\
             Number of users ($K$) &  2 &  Number of RISs ($R$) &  2 \\
             Noise power ($\sigma_n^2$) &  -85 dBm &  Initial learning rate &  $5 \times 10^{-4}$ \\
             Weighting decay &  $2.5 \times 10^{-5}$ &  Epochs &  400 \\
             Batch size &  128 &  Hidden size &  512 \\
            \bottomrule
        \end{tabular}
    }
\end{table}
In this section, numerical simulations are provided to verify the effectiveness of the proposed heterogeneous GNN {  in Table \ref{tab:hyperparams}}.  The BS is assumed to be located at the origin, and two RISs are positioned at coordinates $[30\,m, 25\,m]$ and $[30\,m, -25\,m]$. Additionally, two users are randomly located within a rectangular region defined by the range $[40\,m:50\,m, -25\,m:25\,m]$.
{  According to \cite{akdeniz2014millimeter}, the number of paths $L$ is 3, including one LoS path and $L$-1 NLoS paths. Accordingly, the channel gain for the LoS path is $\beta_{LoS} \sim \mathcal{C N}\left(0,10^{-0.1 \mathrm{PL}(r)}\right)$ where $\mathrm{PL}(r)= \varrho_a + 10 \varrho_b { \log_{10}(r)} + \xi$ with $\xi \sim \mathcal{N}(0,\sigma_{\xi}^2)$. In addition, the channel gain for the NLoS path is $\beta_{NLoS} = 0.01 \beta_{LoS}$ and the channel realizations are obtained by setting $\varrho_{\mathrm{a}}=61.4, \varrho_{\mathrm{b}}=2$, and $\sigma_{\xi}=5.8 \mathrm{~dB}$ \cite{cao2020intelligent}. Without loss of generality, the weights $w_k$ are set to be equal. }
{ 
The proposed heterogeneous GNN is trained and tested separately for each SNR point to evaluate its performance robustly across varying conditions. A total of 110,000 samples, split into 100,000 for training and 10,000 for validation, are utilized. 
}As for the training process, the Adam optimizer \cite{kingma2014adam} is adopted and an NVIDIA RTX $2080$Ti GPU is utilized.  


\subsection{{ Baseline Case Studies}}
{ 
We introduce three case studies to show the significance of selecting a suitable RIS for the users, i.e., considering the scenario when the RIS association scheme is predefined and the heterogeneous GNN just needs to output the beamforming matrix and phase configurations of the RISs. The cases are listed as follows

\begin{itemize}
\item \textbf{Case 1}: All the users are connected to the same RIS, i.e., the system reduces to a single RIS-aided system. 
Therefore, the association matrix $\mathbf{U}$ is reduced to the vector ${\mathbf{u}}^{\rm{(c_1)}}$ and does not need to be optimized. In this case, the message-passing paradigm is illustrated in Section III-B, but without returning ${\mathbf{u}}^{\rm{(c_1)}}$. Therefore, the cascaded channel $ \mathbf{h}_{k}^{(\rm c_1)}$ in the single RIS-aided communication system can be rewritten as
\begin{align}
 \mathbf{h}_{k}^{(\rm c_1)}
 & = u_{k}^{(\rm c_1)}  \mathbf{h}_{1k} {\mathbf{\Phi}}_1^{\rm{(c_1)}} \mathbf{G}_1 \nonumber \\
&= u_{k}^{(\rm c_1)} {\boldsymbol{\theta}}_{1}^{\rm(c_1)} \mathbf{H}_{1k},   
\end{align}
where $\mathbf{H}_{1k} = \mathrm{diag}(\mathbf{h}_{1k}) \mathbf{G}_{1}$ is the input of the heterogeneous GNN. In the single RIS-aided scenario, $\mathbf{u}^{(\rm{c_1})}$ is an all-ones vector, so the channel can be written as $ \mathbf{h}_{1k} =  {\boldsymbol{\theta}}_1^{\rm(c_1)} \mathbf{H}_{1k}$ with $\boldsymbol{\theta}_{1}^{\rm{(c_1)}} = \left[\theta_{1}^{\rm{(c_1)}},\cdots,\theta_{1,M}^{\rm{(c_1)}}\right]$.
Therefore, the WSR maximization problem aims to learn the beamforming matrix $\mathbf{F}^{\rm(c_1)}$ and the phase vector $\boldsymbol{\theta}^{\rm(c_1)}$, which is reformulated as
\begin{subequations}
\begin{align}
\text { P3: }
{\mathop{\max}\limits_{\mathbf{F}^{\rm(c_1)},\boldsymbol{\theta}_1^{\rm(c_1)} }} & \sum_{k = 1}^{K}w_k \rm{log}_2(1+ {\rm{SINR}}_k^{\rm(c_1)}) \\
\text { s.t. } 
& \left\|\mathbf{F}^{\rm(c_1)}\right\|_\mathcal{F}^{2} \leq P_{{  {\max}}} \\
& \left|\boldsymbol{\theta}_{1,m}^{\rm(c_1)}\right| = 1,  \forall m = 1,2, \cdots, M. 
\label{P2}
\end{align}
\end{subequations}
Then, according to the same message passing paradigm and the normalization layers in the orange rectangle in Fig. \ref{structure}, the optimal $\mathbf{F}^{\rm(c_1)}$ and $\boldsymbol{\theta}_1^{\rm(c_1)}$ with the WSR loss function in  (\ref{loss1}).

\item \textbf{Case 2}: All the users are connected to the nearest RIS, then the heterogeneous GNN  outputs $\mathbf{F}^{\rm(c_2)}$ and $\boldsymbol{\theta}_{ i}^{\rm(c_2)}, \forall i \in \mathcal{R}$. Thus, the training process is the same as the pre-training process in the last section and is not illustrated again here.
 
\item \textbf{Case 3}: All the users are connected to the farthest RIS, which is considered as the worst case. Then the heterogeneous GNN outputs $\mathbf{F}^{\rm(c_3)}$ and $\boldsymbol{\theta}_{i}^{\rm(c_3)}, \forall i \in \mathcal{R}$. Similar to Case 2, the phase configuration matrices at multiple RISs need to be optimized, and the cascaded link is expressed as $\mathbf{h}_{k}^{\rm(c_3)}=\sum_{i \in \mathcal{R}} u_{k,i}^{(\rm c_3)} {\boldsymbol{\theta}}_{i}^{\rm(c_3)} \mathbf{H}_{ik} $. Also, the loss function 
is the same as in (\ref{loss1}) by replacing the channel information with $\mathbf{h}_{k}^{\rm(c_3)}$. 
\end{itemize}

}

\subsection{Baseline Algorithms}
{ 
For performance comparison, we consider two typical benchmarks and one deep learning benchmark, including the weighted minimum mean-square error (WMMSE) algorithm with random phases, the AO algorithm for the joint optimization of the BS beamforming matrix at the BS and phase configurations at the RISs, and a DNN algorithm that also employs an encoder-processing-decoder framework to intelligently optimize the BS beamforming matrix at the BS, the phase configurations at the RISs and the RIS association scheme. These methods are illustrated as follows.
}
\begin{itemize}
\item{ {\textbf{Benchmark 1  {  (Random Phase)}}}: The WMMSE \cite{shi2011iteratively} is utilized to optimize the beamforming matrix with a random phase configuration at each RIS, which has been studied extensively in the literature. Then, the iterative updating rule is given as
\begin{align}
\chi_k & =\left(\sum_{i=1}^K\left|\mathbf{h}_k^{\mathrm{H}} \mathbf{f}_i\right|^2+\sigma_0^2\right)^{-1} \mathbf{h}_k^{\mathrm{H}} \mathbf{f}_k, \\
\kappa_k & =\left(1-\chi_k^* \mathbf{h}_k^{\mathrm{H}} \mathbf{f}_k\right)^{-1}, \\
\mathbf{f}_k & =\omega_k \chi_k \kappa_k\left(\lambda \mathbf{I}_M+\sum_{i=1}^K \omega_i\left|\chi_i\right|^2 \kappa_i \mathbf{h}_i \mathbf{h}_i^{\mathrm{H}}\right)^{-1} \mathbf{h}_k,
\end{align}
where $\lambda \geq 0$ is the optimal dual variable for the transmit power constraint. The complexity of the WMMSE algorithm is $\mathcal{O}\left(I_\lambda I_{\mathrm{f}} K N_t^3\right)$, where $I_\lambda$ and $I_{\mathrm{f}}$ are the numbers of iterations for searching $\lambda$ and the three-step updating loop, respectively.
}
 \begin{figure}[ht]
\setlength{\abovecaptionskip}{0pt}
\setlength{\belowcaptionskip}{0pt}
\centering
\includegraphics[width= 0.35\textwidth]{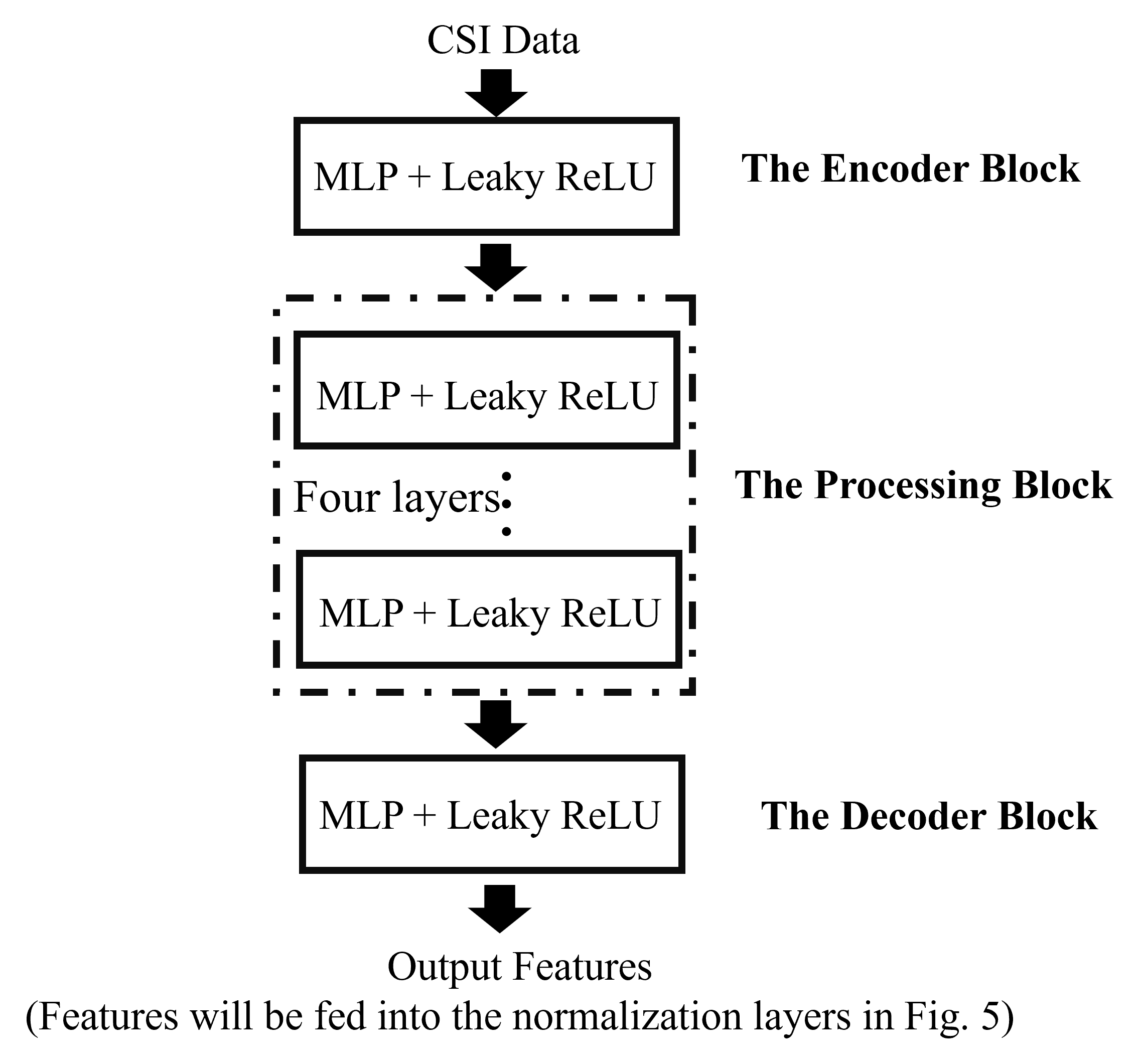}
\DeclareGraphicsExtensions.
\caption{{  Illustration of the { intelligent} DNN benchmark.}}
\label{DNN}
\end{figure}
\item{ {\textbf{Benchmark 2 { (AO)}}}: The AO algorithm is utilized as a performance {  benchmark}, which jointly optimizes the beamforming matrix at the BS and phase configurations at the RISs \cite{Guo2020weighted}.  Based on the AO method, the joint optimization problem is decomposed into two subproblems: one is the conventional beamforming design problem for $\mathbf{F}$ at the BS by using the WMMSE algorithm, and the other is the phase optimization problem for $\mathbf{\Phi}$ according to given optimized beamforming vectors by using the Riemannian conjugate gradient algorithm (RCG) \cite{boumal2014manopt}. Thus, the total complexity of the alternating optimization approach is $\mathcal{O}\left(I_{\mathrm{O}}\left(I_\lambda I_{\mathrm{f}} K N_t^3+I_{\mathrm{R}} K^2 M^2\right)\right)$, where $I_{\mathrm{O}}$ and $I_{\mathrm{R}}$ denote the number of iterations of the outer loop, and the number of iterations of the inner RCG algorithm, respectively.
}

\item{  {\textbf{Benchmark 3 {   (Intelligent DNN)}}}: To validate the advantages of the message-passing scheme for the considered heterogeneous GNN, we compare it with an intelligent DNN benchmark that also employs an encoder-processing-decoder framework. Similar to the proposed heterogeneous GNN, the { intelligent} DNN begins by concatenating all the CSI as inputs. An encoder block with one MLP and a Leaky ReLU layer extracts the features, which are then refined through a processing block that is constituted by four MLPs and Leaky ReLU layers. A decoder block outputs the desired features for problem P1, as shown in Fig. \ref{DNN}. For a fair comparison, the size of the MLP layers of the {  intelligent} DNNs is the same as that of the GNNs.

}

\end{itemize}

\begin{figure*}[htb]
\centering
\includegraphics[width=0.9 \textwidth]{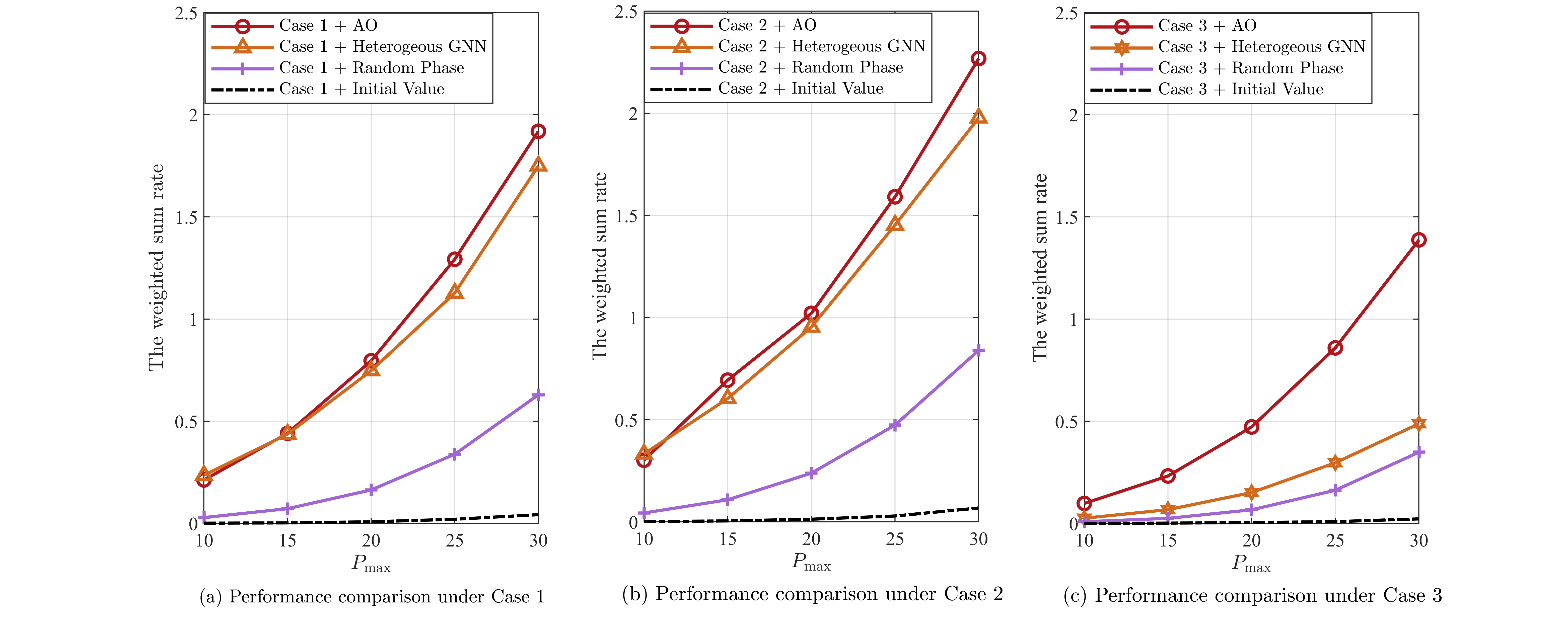}
\caption{The WSR versus $P_{ {\max}}$ { with $N_t = 8$ and $M = 16$. }}
\label{5}
\end{figure*}

\subsection{Performance Versus $P_{ {\max}}$}
  \begin{figure}[ht]
\setlength{\abovecaptionskip}{0pt}
\setlength{\belowcaptionskip}{0pt}
\centering
\includegraphics[width= 0.49\textwidth]{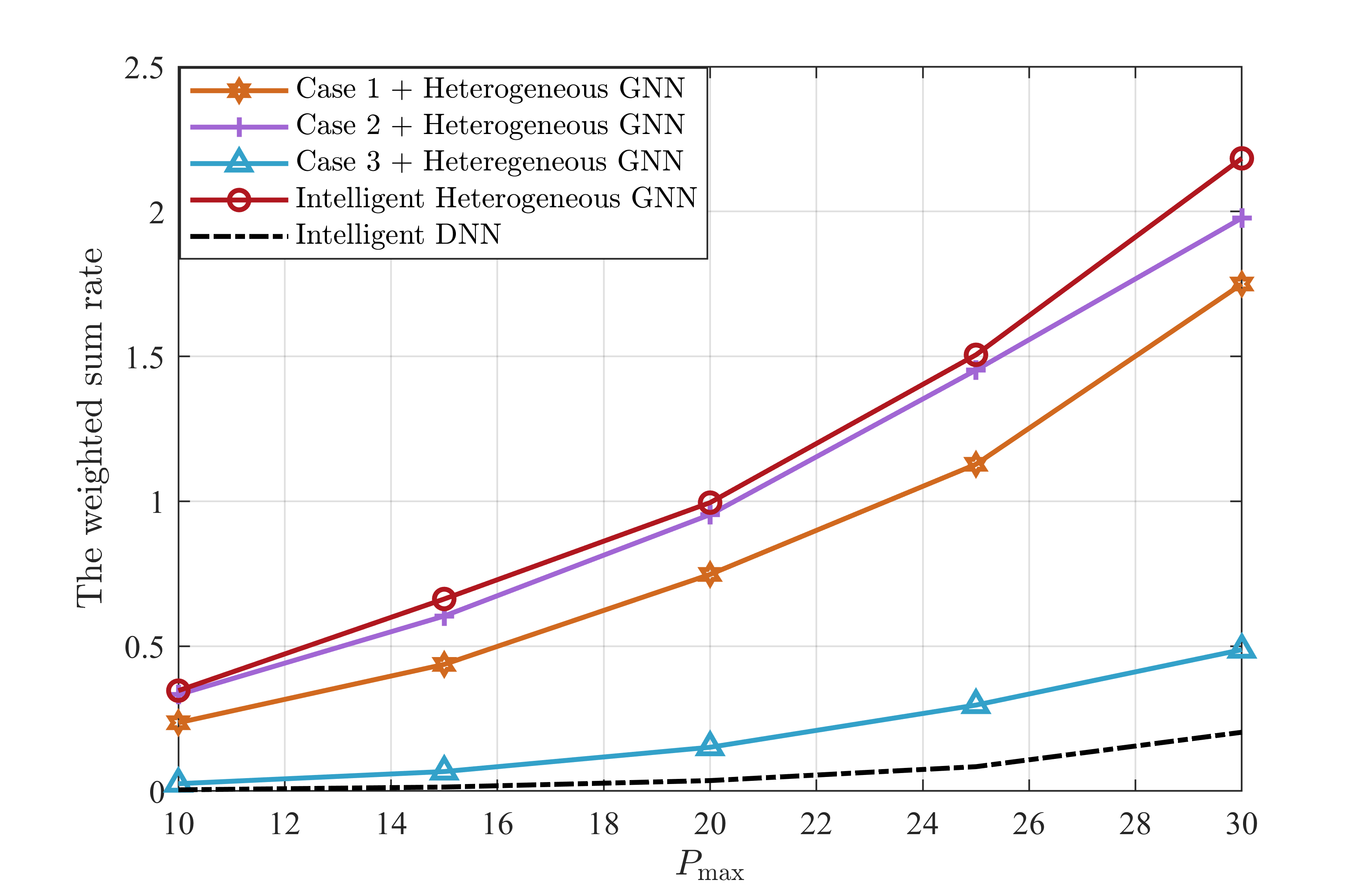}
\DeclareGraphicsExtensions.
\caption{The WSR versus $P_{ {\max}}$ { with $N_t = 8$ and $M = 16$.}}
\label{7}
\end{figure}
{ 
Fig. \ref{5} presents the  WSR as a function of the maximum transmit power \( P_{\max} \) of the BS in a multi-RIS aided system. Specifically, the performance of three different approaches is compared: the AO algorithm and the heterogeneous GNN algorithm (optimizing the beamforming matrix \( \mathbf{F} \) and phase shifts \( \boldsymbol{\theta}_i, \forall i \in \mathcal{R} \)), and WMMSE { with random phase} under three distinct baseline case studies, as shown in Fig. \ref{5}(a)-(c).
From the comparison of these subfigures, it is evident that the WSR increases with rising \( P_{{\max}} \) across all cases. However, different RIS association schemes lead to significant performance variations depending on the proximity of the RIS to the associated users. As far as Case 2 in Fig. \ref{5}(b) is concerned, the WSR consistently outperforms the other case studies. Conversely, when users are connected to RISs located far away as per Case 3,  the performance degrades due to the more severe path loss. Specifically, the RIS association scheme considered in Case 2 offers a notable performance improvement over Case 1 ($18.2\%$) and Case 3 ($63.4\%$), highlighting the importance of selecting an optimal RIS association strategy.
}
  \begin{figure}[ht]
\setlength{\abovecaptionskip}{0pt}
\setlength{\belowcaptionskip}{0pt}
\centering
\includegraphics[width= 0.47\textwidth]{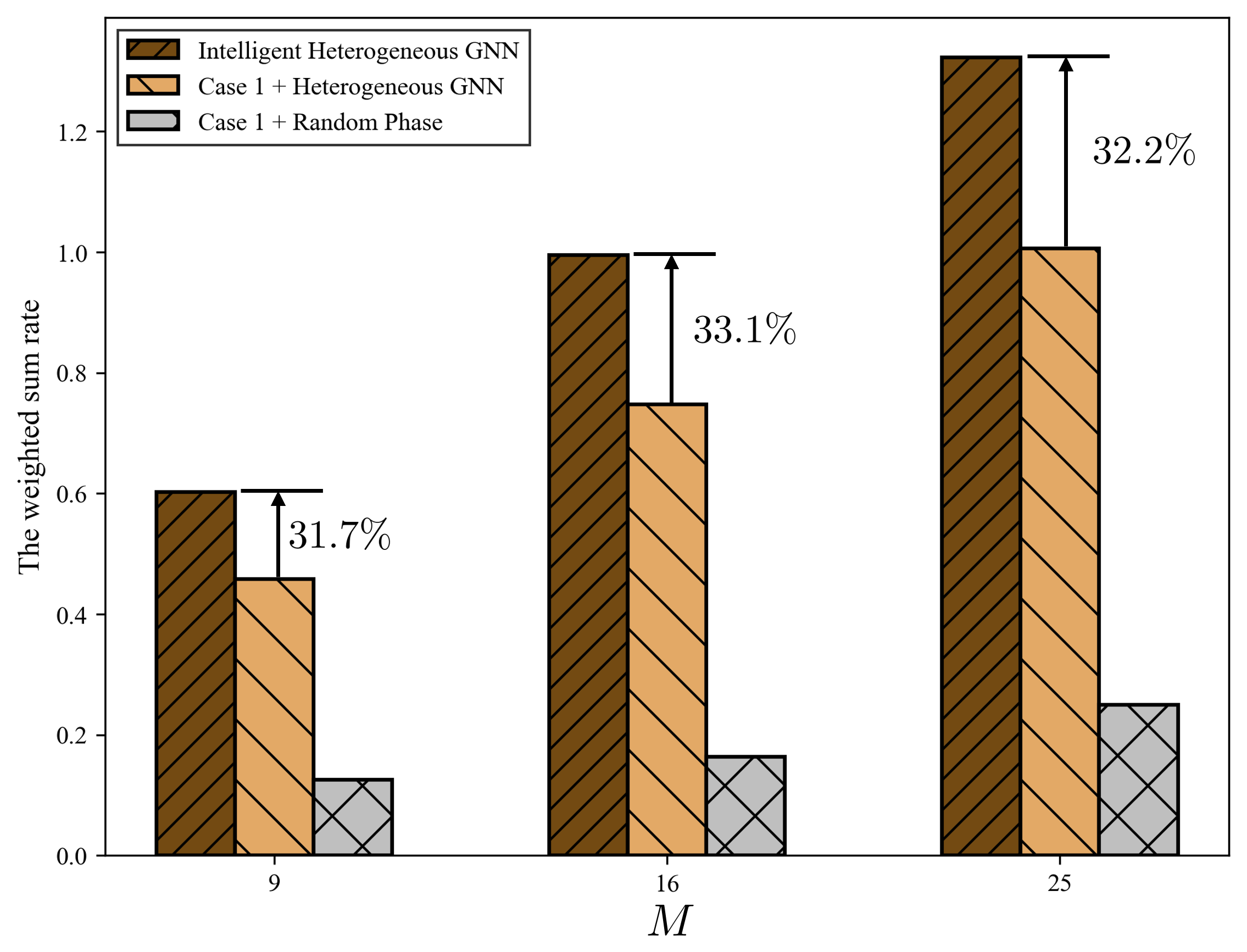}
\DeclareGraphicsExtensions.
\caption{The WSR versus $M$ with $P_{ {\max}} = 20$ dBm {  and $N_t$ = 8}.}
\label{8}
\end{figure}
{ 

Fig. \ref{7} presents a performance comparison across different network architectures, including the combination of three case studies with the heterogeneous GNN, the intelligent heterogeneous GNN, and the intelligent DNN. Here, `intelligent' indicates that the association scheme \( \mathbf{U} \) is learned by the neural network simultaneously with the beamforming matrix \( \mathbf{F} \) and phase configurations \( \boldsymbol{\theta}_i, \forall i \in \mathcal{R} \), as opposed to the two-step methods utilized by the baseline case studies.
As shown in Fig. \ref{7}, the intelligent heterogeneous GNN outperforms the other architectures, indicating that this intelligent learning approach yields significant improvements due to the coupled { optimization of all variables. Specifically, the intelligent heterogeneous GNN demonstrates a $10.7\%$ improvement compared with Case 2 + heterogeneous GNN, and the intelligent heterogeneous GNN is shown to promote the association of each user to its nearest RIS {  in light of the multiplicative path loss effects discussed in \cite{basar2021present}}, as guided by the CE loss function in (\ref{loss2}) during training.
 Additionally, it is evident that the intelligent DNN performs the worst, primarily due to its inability to effectively account for the underlying network structure.}

}

\subsection{Performance Versus $M$}
  \begin{figure}[ht]
\setlength{\abovecaptionskip}{0pt}
\setlength{\belowcaptionskip}{0pt}
\centering
\includegraphics[width= 0.47\textwidth]{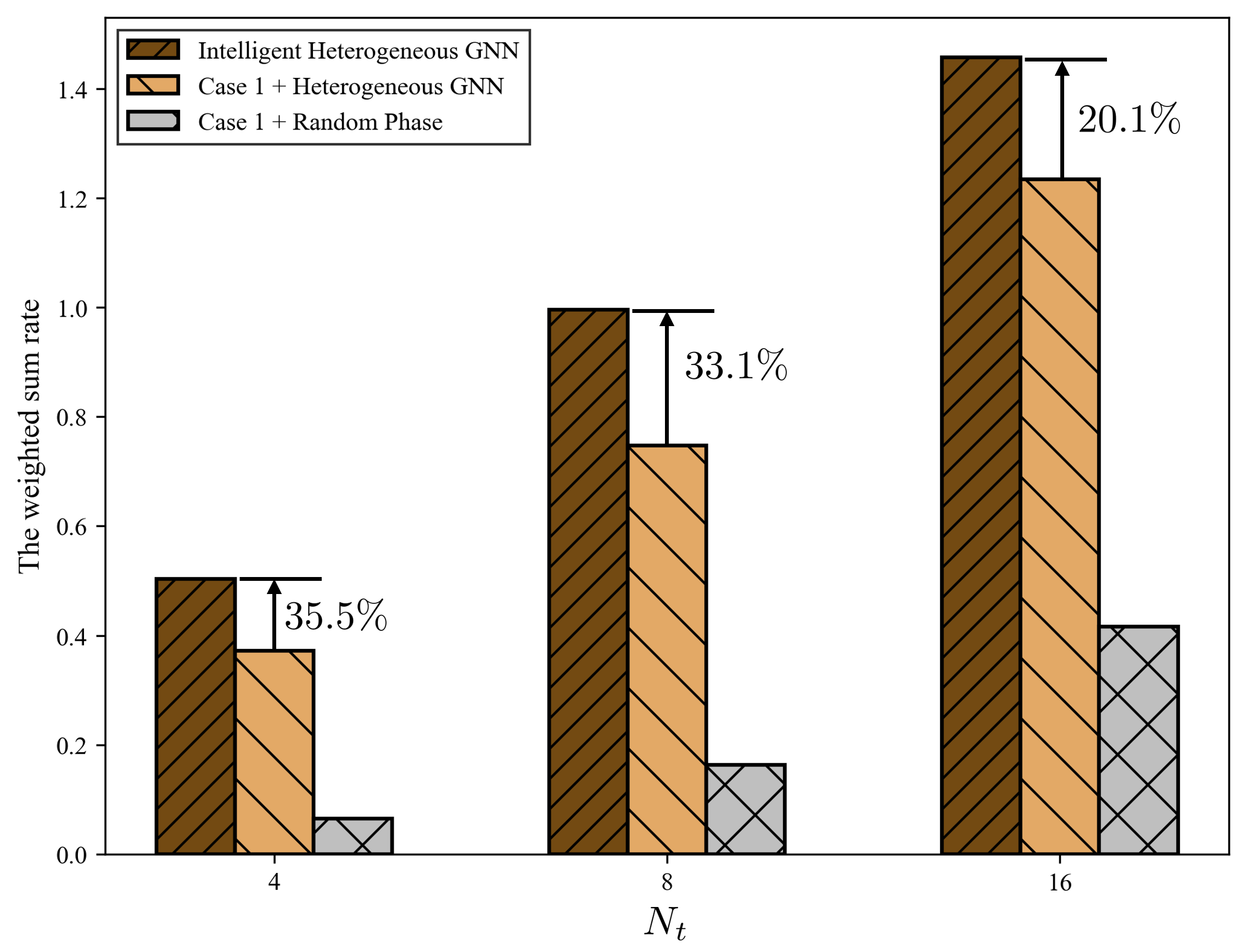}
\DeclareGraphicsExtensions.
\caption{The WSR versus $N_t$ with $P_{ {\max}} = 20$ dBm {  and $M = 16$}.}
\label{9}
\end{figure}
  \begin{table}[!h]
  \caption{\textbf{COMPUTATION TIME COMPARISON FOR DIFFERENT ALGORITHMS { WITH $P_{{\max}} = 20$ dBm {  AND $M = 16$} 
  [SEC]}}}
 \centering
 \scalebox{0.85}
 {
\begin{tabular}{c c c c c}
\hline
\specialrule{0em}{1pt}{1pt}
\hline
 & Algorithm &$N_t = 4$ &  $N_t = 8$ & $N_t = 16$ \\
 \hline
\multirow{2}{*}{CPU} &{  Case 2 + AO} & $1.2381$ & $1.4633$ & $2.0497$ \\
\cmidrule(l){2-5}
&{  Intelligent Heterogeneous GNN} & $4.984${ E}${-2}$ &  $5.248${ E}${-2}$ & $5.351${ E}${-2}$ \\
\midrule
\multirow{2}{*}{GPU} & {  Case 2 + Heterogeneous GNN} & $2.297${ E}${-2}$ & $2.295${ E}${-2}$ & $2.389${ E}${-2}$ \\
\cmidrule(l){2-5}
&{  Intelligent Heterogeneous GNN} & $2.244${ E${-2}$} &  $2.250${ E${-2}$} & $2.374${ $E{-2}$} \\
\hline
\specialrule{0em}{1pt}{1pt}
\hline
\end{tabular}}
\label{complexity}
\end{table}
 Fig. \ref{8} shows the WSR versus the number of reflecting elements at each RIS $M$. We see that the WSR improves monotonically as $M$ increases. Besides this, the proposed heterogeneous GNN structure can tackle the optimization problem P1 for a different number of reflecting elements {\color {black} ($M = 9, 16, 25$)}, showcasing the generality and effectiveness of the proposed architecture. Then, comparing the results of {  random phase} under Case 1, we see that $M$ has little influence on the performance because the {  random phase} does not optimize the phase configurations of the RISs. 
In general, the RIS association scheme results in a performance gain of about { 32.3$\%$} for all values of $M$.
 
 \subsection{Performance Versus $N_t$}

 { 

Fig. \ref{9} illustrates the WSR versus the total number of antennas \( N_t \). It is evident that the WSR increases with the number of antennas and the proposed { intelligent} heterogeneous GNN consistently illustrates strong performance across different values of \( N_t \). 
Moreover, the results show that selecting an appropriate RIS for each user significantly enhances performance. The performance improvement ranges from approximately 20\% to 35\% depending on the number of antennas, highlighting the effectiveness of the intelligent heterogeneous GNN in selecting the suitable association scheme.
 }
 \subsection{Complexity and Convergence of the Heterogeneous GNN }

\begin{figure}[ht]
\setlength{\abovecaptionskip}{0pt}
\setlength{\belowcaptionskip}{0pt}
\centering
\includegraphics[width= 0.49\textwidth]{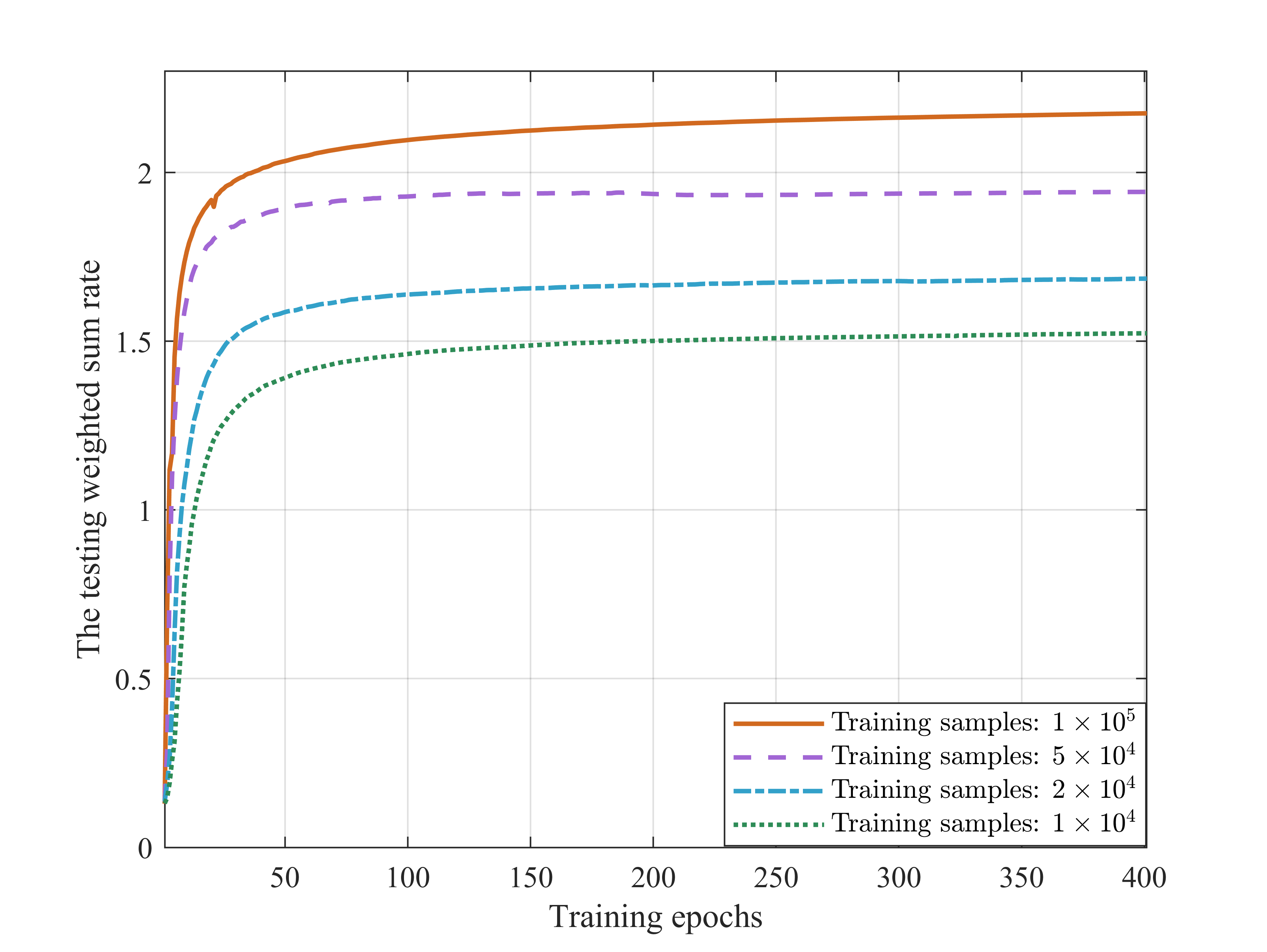}
\DeclareGraphicsExtensions.
\caption{The testing WSR versus training epochs { with $N_t = 8$, $M = 16$ and $P_{\max} = 30$ dBm.}}
\label{epoch}
\end{figure}

The complexity of different algorithms and the convergence of the proposed { intelligent} heterogeneous GNN are illustrated in Table \ref{complexity} and Fig. \ref{epoch}. In Table \ref{complexity}, the running time of different methods for different values of $N_t$ are reported.

{ 
We executed the AO algorithm considering Case 2 and the { intelligent} heterogeneous GNN on an 11th Gen Intel Core i7-1165G7 @ 2.80GHz CPU, as well as on an NVIDIA RTX 2080Ti GPU. Running times were evaluated across $128$ problem instances, corresponding to the batch size utilized during training.
The { intelligent} heterogeneous GNN run on the GPU  utilizes parallel computing, unlike the AO algorithm, which is sequential by nature. As \( N_t \) increases, the running {  time of the AO algorithm increases significantly. 
Notably, the computation time of the GPU for the { intelligent} heterogeneous GNN, including the additional loss function for user-RIS association learning, is comparable to the heterogeneous GNN under Case 2, indicating minimal impact on time complexity. On the CPU, the running time of the GNN is about twice that of the GPU but still only $2.5\%$ of the time required by the AO algorithm. {  Additionally, the { intelligent} heterogeneous GNN comprises roughly 5.8 million parameters, a moderate count that underscores the network's lower complexity and supports the running time analysis as a metric for validation.} In conclusion, the comparison of running times underscores the advantage of the intelligent heterogeneous GNN, which achieves performance comparable to the AO algorithm under Case 2 while significantly reducing the computation time.}}

Then we analyze the test results. Specifically,  we plot the testing WSR evaluated on different numbers of training samples against the training epochs, to evaluate the convergence of the proposed { intelligent} heterogeneous GNN in Fig. \ref{epoch}. The { intelligent} heterogeneous GNN is trained for $400$ epochs under different numbers of training samples. We see that the WSR initially increases sharply with the training epoch but it then converges to a stable value after $50$ training epochs. Therefore, a few training epochs are sufficient to achieve satisfactory performance. If $1 \times 10^5$ training samples are considered, for instance, greater than $93\%$ of the WSR is achieved with only $50$ training epochs.

 \subsection{ Performance Versus the Training Parameters}
 
 \begin{figure}[ht]
\setlength{\abovecaptionskip}{0pt}
\setlength{\belowcaptionskip}{0pt}
\centering
\includegraphics[width= 0.49\textwidth]{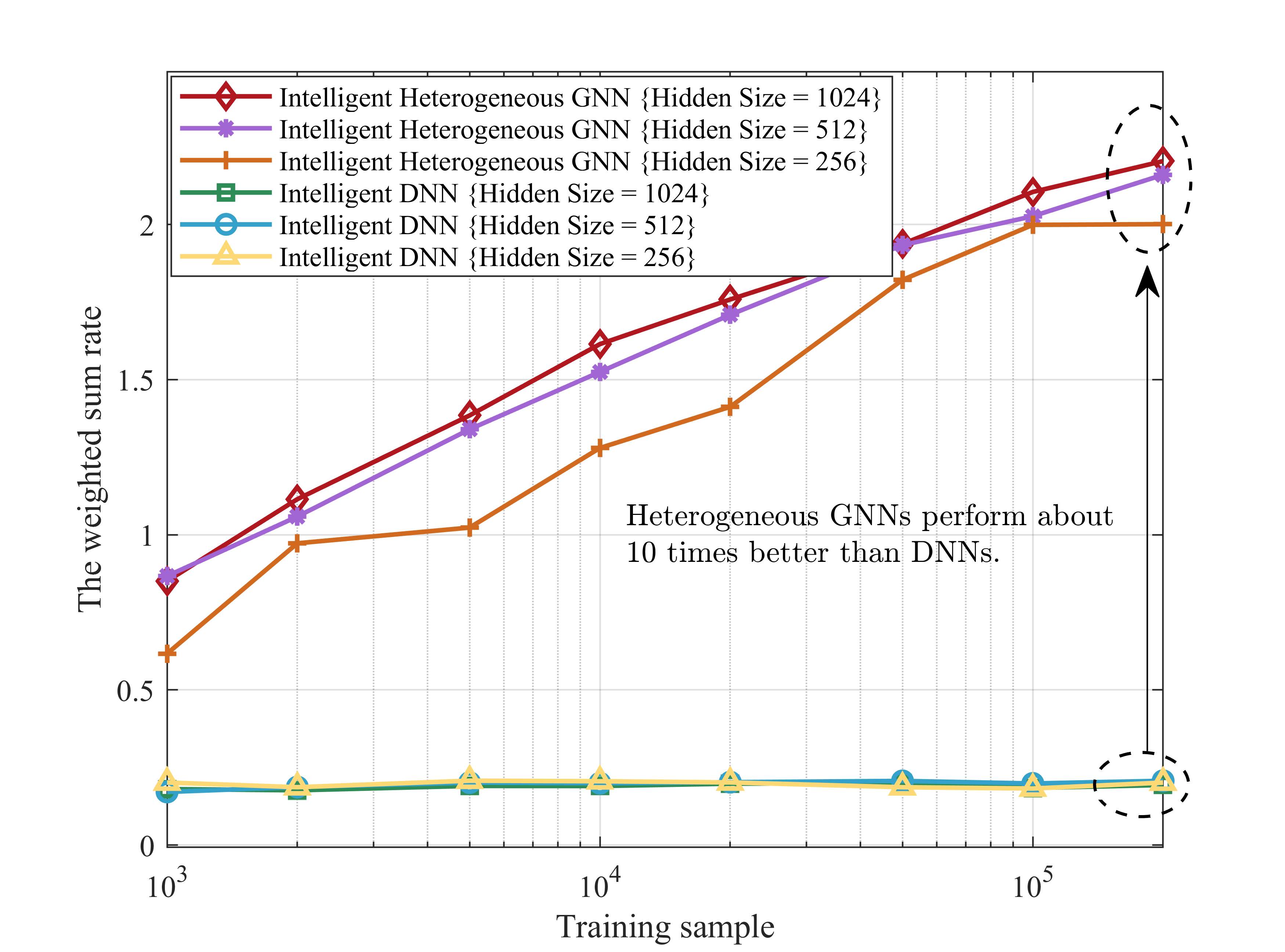}
\DeclareGraphicsExtensions.
\caption{The WSR versus the training samples { with $N_t = 8$, $M = 16$ and $P_{\max} = 30$ dBm.}}
\label{11}
\end{figure}

{ 
Next, we present the WSR performance as a function of the number of training samples and analyze the impact of different model sizes for the proposed intelligent heterogeneous GNNs and intelligent DNNs.
{  In contrast to the baseline case studies that utilize a two-step method---initially determining the association scheme \(\mathbf{U}\) and subsequently calculating the beamforming matrix \(\mathbf{F}\) and phase configurations \(\mathbf{\theta}_i, \forall i \in \mathcal{R}\), both intelligent neural networks output \(\mathbf{U}\), \(\mathbf{F}\), and \(\mathbf{\theta}_i, \forall i \in \mathcal{R}\) simultaneously.}
As shown in Fig. \ref{11}, the performance of the intelligent GNN improves with the number of training samples and the size of the layers. Notably, the difference in performance between the intelligent heterogeneous GNNs with hidden sizes of $512$ and $1024$ is minor, indicating that a hidden size of $512$ is sufficient for the considered optimization problem.
The intelligent heterogeneous GNNs significantly outperform the intelligent DNNs. Assuming \(2 \times 10^{5}\) training samples, for example, the intelligent GNNs perform $10$ times better than the DNNs. Conversely, increasing the hidden sizes or training samples does not enhance the performance of DNNs. This is because the RIS association scheme strengthens the impact of structural information on system optimization, and DNNs are incapable of capturing the underlying network structure. Unlike the DNNs, which require large amounts of training data, the proposed intelligent GNNs provide better results with fewer samples. Also, their performance can be improved as the number of training samples increases.
 }

\section{Conclusion}

In this work, we proposed a heterogeneous GNN structure to solve the NP-hard rate optimization problem in multi-RIS aided mmWave communication systems. The RIS association scheme was shown to significantly improve the system performance while increasing the optimization complexity. By mapping the structure of the wireless environment into a heterogeneous GNN, the optimal beamforming matrices and RIS phase configurations were obtained to maximize the WSR in the considered multi-RIS aided communication system.
Extensive simulation results were performed to validate the performance of the proposed GNN. Specifically, it was obtained that the proposed heterogeneous GNN outperformed conventional DNNs by a factor of 10, as the latter is not capable of capturing the underlying structure of the wireless environment.
Moreover, simulation results showed that the RIS association scheme can improve the system performance by about 30$\%$.
{ 
Potential extensions of this study could involve accommodating a variable number of users, assessing deployment costs, and evaluating interference from multiple non-associated RISs. Additionally, exploring the use of soft labels, such as normalized inverse distance distributions instead of traditional binary one-hot coding, may offer further enhancements in performance.
}

\bibliographystyle{IEEEtran}
\bibliography{IEEEabrv,myref}

\end{document}